\documentclass[aps,prd,superscriptaddress,nofootinbib,preprint]{revtex4}
\usepackage{ulem}
\usepackage{color}
\usepackage{graphicx}
\usepackage{multirow}
\usepackage{bm}

\begin{document}

\title{Getting the Most from Detection of Galactic Supernova Neutrinos in Future Large Liquid-Scintillator Detectors}

\author{Jia-Shu Lu}

\email{lujiashu@ihep.ac.cn}

\affiliation{Institute of High Energy Physics, Chinese Academy of Sciences, Beijing 100049, China}

\author{Yu-Feng Li}

\email{liyufeng@ihep.ac.cn}

\affiliation{Institute of High Energy Physics, Chinese Academy of Sciences, Beijing 100049, China}

\author{Shun Zhou}

\email{zhoush@ihep.ac.cn}

\affiliation{Institute of High Energy Physics, Chinese Academy of Sciences, Beijing 100049, China}
\affiliation{Center for High Energy Physics, Peking University, Beijing 100871, China}

\begin{abstract}
Future large {liquid-}scintillator detectors can be implemented to observe neutrinos from a core-collapse supernova (SN) in our galaxy in various reaction channels: (1) The inverse beta decay $\overline{\nu}^{}_e + p \to n + e^+$; (2) The elastic neutrino-proton scattering $\nu + p \to \nu + p$\,; (3) The elastic neutrino-electron scattering $\nu + e^- \to \nu + e^-$; (4) The charged-current $\nu^{}_e$ interaction $\nu^{}_e + {^{12}}{\rm C} \to e^- + {^{12}}{\rm N}$; (5) The charged-current $\overline{\nu}^{}_e$ interaction $\overline{\nu}^{}_e + {^{12}}{\rm C} \to e^+ + {^{12}}{\rm B}$; (6) The neutral-current interaction $\nu + {^{12}}{\rm C} \to \nu + {^{12}}{\rm C}^*$. {The less abundant $^{13}{\rm C}$ atoms in the liquid scintillator are also considered as a target, and both the charged-current interaction $\nu^{}_e + {^{13}}{\rm C} \to e^- + {^{13}}{\rm N}$ and the neutral-current interaction $\nu + {^{13}}{\rm C} \to \nu + {^{13}}{\rm C}^*$ are taken into account.} In this work, we show for the first time that a global analysis of all these channels at a single {liquid-}scintillator detector, such as Jiangmen Underground Neutrino Observatory (JUNO), is very important to test the average-energy hierarchy of SN neutrinos and how the total energy is partitioned among neutrino flavors. In addition, the dominant channels for reconstructing neutrino spectra and the impact of other channels are discussed in great detail.
\end{abstract}

\maketitle

\section{Introduction}

How massive stars eventually end their lives remains an open question in astrophysics and astronomy. In particular, a star of mass above eight solar masses or so is expected to experience a core collapse under its own gravity, and then a violent explosion~\cite{Bethe:1990mw}. In the paradigm of neutrino-driven explosion, it is neutrinos that restore most of the gravitational binding energy released in the core collapse and deposit part of it to the surroundings, reviving the halted shock wave and leading to a successful explosion~\cite{Wilson:1985,Bethe:1984ux}. Therefore, the detection of neutrino signals and reconstruction of neutrino energy spectra are of crucial importance to verify the neutrino-driven paradigm, and finally reveal the true mechanism for core-collapse SN explosions~\cite{Mirizzi:2015eza}.

The interest in SN neutrino detection has recently been stimulated by the great progress in neutrino oscillation experiments, for which large water-Cherenkov and {liquid-}scintillator detectors are under construction or will be built in the near future. For a galactic core-collapse SN at a typical distance of $10~{\rm kpc}$, the Super-Kamiokande detector (SK) is capable of collecting about $10^4$ neutrino events mainly in the inverse beta decay channel $\overline{\nu}^{}_e + p \to e^+ + n$ (IBD). In Ref.~\cite{Laha:2013hva}, it has been demonstrated that the SK with a gadolinium-loaded water Cherenkov detector opens a possibility to determine, with a precision of $20\%$, the total and average energy of $\nu^{}_e$ via the elastic neutrino-electron scattering $\nu^{}_e + e^- \to \nu^{}_e + e^-$ ($e$ES). The precision could be further improved by a factor of five in the future Hyper-Kamiokande with dissolved gadolinium~\cite{Laha:2013hva}. In addition, the JUNO experiment is designed to determine neutrino mass ordering by precisely measuring the spectrum of reactor antineutrinos, with a 20 kiloton liquid-scintillator detector~\cite{Li:2013zyd,An:2015jdp}. Undoubtedly, JUNO will also serve as a powerful detector for galactic SN neutrinos. For instance, we have found in Ref.~\cite{Lu:2014zma} that JUNO is even better than SK in constraining the absolute neutrino masses through the time-delay effects. See, Ref.~\cite{Rossi-Torres:2015rla}, for the study for a future liquid-argon detector.

Future large {liquid-}scintillator detectors, such as JUNO~\cite{An:2015jdp}, RENO-50~\cite{Kim:2014rfa} and LENA~\cite{Wurm:2011zn}, will have several advantages for the detection of SN neutrinos. First of all, the threshold of visible energy in a {liquid-}scintillator detector could be rather low, since it is only limited by the intrinsic radioactive backgrounds of the {liquid} scintillator itself. If the abundance of $^{14}{\rm C}$ can be controlled at the level already achieved in the Borexino experiment, the energy threshold will be as low as $0.2~{\rm MeV}$~\cite{Alimonti:2000xc}. In this case, the elastic neutrino-proton scattering $\nu + p \to \nu + p$ ($p$ES) becomes very important, giving rise to a large number of events in a channel other than IBD. Second, the carbon nuclei in the {liquid} scintillator serve as an invaluable target for SN neutrino detection. In particular, the SN $\nu^{}_e$ is detectable via the charged-current interaction $\nu^{}_e + {^{12}}{\rm C} \to e^- + {^{12}}{\rm N}$ (${^{12}}{\rm N}$-CC) in addition to the elastic scattering off electrons and protons. {Although the natural abundance of $^{13}{\rm C}$ on Earth is small (i.e., about $1.1\%$), the charged-current interaction $\nu^{}_e + {^{13}}{\rm C} \to e^- + {^{13}}{\rm N}$ (${^{13}}{\rm N}$-CC) and the neutral-current interaction $\nu + {^{13}}{\rm C} \to \nu + {^{13}}{\rm C}^*$ (${^{13}}{\rm C}$-NC) have been proposed as promising channels to detect solar ${^8}{\rm B}$ neutrinos~\cite{Arafune:1988hx,Fukugita:1989wv,Suzuki:2012aa,Mollenberg:2014mfa}. These two processes may be important in a massive liquid-scintillator detector and are relevant for the SN neutrino detection.} Third, it is in principle possible to distinguish between protons and photons/electrons/positrons in a {liquid-}scintillator detector with the pulse shape discrimination, implying a remarkable reduction of background in each channel. Therefore, the {liquid-}scintillator detectors are able to provide us with more information about neutrino flavors and energy spectra. The event spectra of SN neutrinos in highly pure {liquid-}scintillator detectors have been studied in the literature~\cite{Lujan-Peschard:2014lta,An:2015jdp}. In Ref.~\cite{Laha:2014yua}, the combined analysis of two dominant channels $e$ES and ${^{12}}{\rm N}$-CC is carried out to probe the total and average energies of $\nu^{}_e$, as a counterpart study for the SK in Ref.~\cite{Laha:2013hva}.

In this paper, we perform for the first time a global analysis of main detection channels of SN neutrinos at a large {liquid-}scintillator detector. The prospects for SN $\nu^{}_e$, $\overline{\nu}^{}_e$ and $\nu^{}_x$ detection are examined with quantitative analyses, where $\nu^{}_x$ collectively stands for $\nu^{}_\mu$, $\nu^{}_\tau$ or their antiparticles. The experimental sensitivities to total and average energies of each neutrino flavor are explored. First, we ignore neutrino flavor oscillations and concentrate on how well the average and total energies of each neutrino flavor can be determined. Then, neutrino oscillations with matter effects in the supernova envelope and the impact of neutrino mass ordering are taken into account. We proceed to reconstruct the total and average neutrino energy from the global analysis, and further test how the energy is distributed among neutrino flavors. The hypothesis of energy equipartition, which has been taken for granted in previous studies of SN neutrinos, is critically studied.

The remaining part of our paper is organized as follows. Sec. II is devoted to a summary of SN neutrino fluxes and the detection channels in future large {liquid-}scintillator detectors. Then, in Sec. III, numerical simulations are carried out for the JUNO detector to determine the total energies $(E^{\rm tot}_{\nu_e}, E^{\rm tot}_{\bar{\nu}_e}, E^{\rm tot}_{\nu_x})$ and average energies $(\langle E^{}_{\nu_e}\rangle, \langle E^{}_{\bar{\nu}_e}\rangle, \langle E^{}_{\nu_x}\rangle)$ of SN neutrinos in three different species $\nu^{}_e$, $\overline{\nu}^{}_e$ and $\nu^{}_x$. In order to clearly see the experimental sensitivities, we temporarily ignore the flavor conversions of SN neutrinos, and concentrate on the impact of {combining} different signal channels. In Sec. IV, the test of energy-equipartition hypothesis is performed in the realistic case by including the flavor conversions in both neutrino mass orderings. Finally, we summarize our main conclusions in Sec. V{, and the details of the $\chi^2$ functions implemented in our calculations are given in the Appendix.}

\section{Supernova Neutrinos at Liquid-Scintillator Detectors}

\subsection{Neutrino Spectra}

For a core-collapse SN, the total gravitational binding energy is about $3 \times 10^{53}~{\rm erg}$, which is mostly carried away by neutrinos and antineutrinos of all three flavors in about ten seconds. According to a detailed Monte-Carlo study of neutrino spectra formation by considering the most relevant microscopic processes, the time-integrated neutrino spectrum can be well described by three parameters: the total energy $E^{\rm tot}_\alpha$, the average energy $\langle E^{}_\alpha \rangle $ and the spectral index $\gamma_\alpha$. In this parametrization, the neutrino fluences are given by~\cite{Keil:2002in}
\begin{equation}
F^0_\alpha(E) = \frac{1}{4\pi r^2} \frac{E^{\rm tot}_\alpha}{\langle E^{}_\alpha \rangle} \frac{(1+\gamma^{}_\alpha)^{1+\gamma^{}_\alpha}}{\Gamma(1+\gamma^{}_\alpha)} \left(\frac{E}{\langle E^{}_\alpha \rangle} \right)^{\gamma^{}_\alpha} \exp\left[-(1+\gamma^{}_\alpha)\frac{E}{\langle E_\alpha \rangle}\right] \; ,
\end{equation}
where $\alpha$ stands for $\nu^{}_e$, $\overline{\nu}^{}_e$ and $\nu^{}_x$, and $r$ for the radius where the neutrino fluxes are evaluated. Usually the total energy is assumed to be equally distributed among all neutrinos and antineutrinos, namely, $E^{\rm tot}_{\nu_e} = E^{\rm tot}_{\bar{\nu}_e} = E^{\rm tot}_{\nu_x} = 5 \times 10^{52}~{\rm erg}$. In our discussions, we also consider the situation when this assumption is relaxed.

\subsection{Detection Channels}

\begin{table}[!t]
\centering
\begin{tabular}{ccccccccc}
\hline \hline
\multicolumn{1}{c}{\multirow {2}{*}{Channel}} & \multicolumn{1}{c}{} & \multicolumn{1}{c}{\multirow {2}{*}{Type}} & \multicolumn{1}{c}{} & \multicolumn{5}{c}{Number of SN Neutrino Events at JUNO} \\
\cline{5-9} \multicolumn{1}{c}{} & \multicolumn{1}{c}{} & \multicolumn{1}{c}{} & \multicolumn{1}{c}{} & \multicolumn{1}{c}{No Oscillations} & \multicolumn{1}{c}{} & \multicolumn{1}{c}{Normal Ordering} & \multicolumn{1}{c}{} & \multicolumn{1}{c}{Inverted Ordering} \\
\hline
\multicolumn{1}{l}{$\overline{\nu}_e + p \to e^+ + n$} & \multicolumn{1}{c}{} & \multicolumn{1}{c}{CC} & \multicolumn{1}{c}{} & \multicolumn{1}{c}{$4573$} & \multicolumn{1}{c}{} & \multicolumn{1}{c}{$4775$} & \multicolumn{1}{c}{} & \multicolumn{1}{c}{$5185$} \\
\cline{1-9} \multicolumn{1}{l}{\multirow {4}{*}{$\nu + p \to \nu + p$}} & \multicolumn{1}{c}{} & \multicolumn{1}{c}{\multirow {4}{*}{ES}} & \multicolumn{1}{c}{} & \multicolumn{1}{c}{$1578$} & \multicolumn{1}{c}{} & \multicolumn{1}{c}{$1578$} & \multicolumn{1}{c}{} & \multicolumn{1}{c}{$1578$} \\
\cline{5-9}
\multicolumn{1}{l}{} & \multicolumn{1}{c}{} & \multicolumn{1}{c}{} & \multicolumn{1}{c}{$\nu^{}_e$} & \multicolumn{1}{c}{$107$} & \multicolumn{1}{c}{} & \multicolumn{1}{c}{$354$} & \multicolumn{1}{c}{} & \multicolumn{1}{c}{$278$} \\
\multicolumn{1}{l}{} & \multicolumn{1}{c}{} & \multicolumn{1}{c}{} & \multicolumn{1}{c}{$\overline{\nu}^{}_e$} & \multicolumn{1}{c}{$179$} & \multicolumn{1}{c}{} & \multicolumn{1}{c}{$214$} & \multicolumn{1}{c}{} & \multicolumn{1}{c}{$292$} \\
\multicolumn{1}{l}{} & \multicolumn{1}{c}{} & \multicolumn{1}{c}{} & \multicolumn{1}{c}{$\nu^{}_x$} & \multicolumn{1}{c}{$1292$} & \multicolumn{1}{c}{} & \multicolumn{1}{c}{$1010$} & \multicolumn{1}{c}{} & \multicolumn{1}{c}{$1008$} \\
\hline
\multicolumn{1}{l}{\multirow {4}{*}{$\nu^{}_e + e \to \nu^{}_e + e$}} & \multicolumn{1}{c}{} & \multicolumn{1}{c}{\multirow {4}{*}{ES}} & \multicolumn{1}{c}{} & \multicolumn{1}{c}{$314$} & \multicolumn{1}{c}{} & \multicolumn{1}{c}{$316$} & \multicolumn{1}{c}{} & \multicolumn{1}{c}{$316$} \\
\cline{5-9}
\multicolumn{1}{l}{} & \multicolumn{1}{c}{} & \multicolumn{1}{c}{} & \multicolumn{1}{c}{$\nu^{}_e$} & \multicolumn{1}{c}{$157$} & \multicolumn{1}{c}{} & \multicolumn{1}{c}{$159$} & \multicolumn{1}{c}{} & \multicolumn{1}{c}{$158$} \\
\multicolumn{1}{l}{} & \multicolumn{1}{c}{} & \multicolumn{1}{c}{} & \multicolumn{1}{c}{$\overline{\nu}^{}_e$} & \multicolumn{1}{c}{$61$} & \multicolumn{1}{c}{} & \multicolumn{1}{c}{$61$} & \multicolumn{1}{c}{} & \multicolumn{1}{c}{$62$} \\
\multicolumn{1}{l}{} & \multicolumn{1}{c}{} & \multicolumn{1}{c}{} & \multicolumn{1}{c}{$\nu^{}_x$} & \multicolumn{1}{c}{$96$} & \multicolumn{1}{c}{} & \multicolumn{1}{c}{$96$} & \multicolumn{1}{c}{} & \multicolumn{1}{c}{$96$} \\
\hline
\multicolumn{1}{l}{$\nu_e + {^{12}}{\rm C} \to e^- + {^{12}}{\rm N}$} & \multicolumn{1}{c}{} & \multicolumn{1}{c}{CC} & \multicolumn{1}{c}{} & \multicolumn{1}{c}{$43$} & \multicolumn{1}{c}{} & \multicolumn{1}{c}{$134$} & \multicolumn{1}{c}{} & \multicolumn{1}{c}{$106$} \\
\hline
\multicolumn{1}{l}{$\overline{\nu}_e + {^{12}}{\rm C} \to e^+ + {^{12}}{\rm B}$} & \multicolumn{1}{c}{} & \multicolumn{1}{c}{CC} & \multicolumn{1}{c}{} & \multicolumn{1}{c}{$86$} & \multicolumn{1}{c}{} & \multicolumn{1}{c}{$98$} & \multicolumn{1}{c}{} & \multicolumn{1}{c}{$126$} \\
\hline
\multicolumn{1}{l}{\multirow {4}{*}{$\nu + {^{12}}{\rm C} \to \nu + {^{12}}{\rm C}^*$}} & \multicolumn{1}{c}{} & \multicolumn{1}{c}{\multirow {4}{*}{NC}} & \multicolumn{1}{c}{} & \multicolumn{1}{c}{$352$} & \multicolumn{1}{c}{} & \multicolumn{1}{c}{$352$} & \multicolumn{1}{c}{} & \multicolumn{1}{c}{$352$} \\
\cline{5-9}
\multicolumn{1}{l}{} & \multicolumn{1}{c}{} & \multicolumn{1}{c}{} & \multicolumn{1}{c}{$\nu^{}_e$} & \multicolumn{1}{c}{$27$} & \multicolumn{1}{c}{} & \multicolumn{1}{c}{$76$} & \multicolumn{1}{c}{} & \multicolumn{1}{c}{$61$} \\
\multicolumn{1}{l}{} & \multicolumn{1}{c}{} & \multicolumn{1}{c}{} & \multicolumn{1}{c}{$\overline{\nu}^{}_e$} & \multicolumn{1}{c}{$43$} & \multicolumn{1}{c}{} & \multicolumn{1}{c}{$50$} & \multicolumn{1}{c}{} & \multicolumn{1}{c}{$65$} \\
\multicolumn{1}{l}{} & \multicolumn{1}{c}{} & \multicolumn{1}{c}{} & \multicolumn{1}{c}{$\nu^{}_x$} & \multicolumn{1}{c}{$282$} & \multicolumn{1}{c}{} & \multicolumn{1}{c}{$226$} & \multicolumn{1}{c}{} & \multicolumn{1}{c}{$226$} \\
\hline
\multicolumn{1}{l}{$\nu_e + {^{13}}{\rm C} \to e^- + {^{13}}{\rm N}$} & \multicolumn{1}{c}{} & \multicolumn{1}{c}{CC} & \multicolumn{1}{c}{} & \multicolumn{1}{c}{$19$} & \multicolumn{1}{c}{} & \multicolumn{1}{c}{$29$} & \multicolumn{1}{c}{} & \multicolumn{1}{c}{$26$} \\
\cline{1-9}\multicolumn{1}{l}{\multirow {4}{*}{$\nu + {^{13}}{\rm C} \to \nu + {^{13}}{\rm C}^*$}} & \multicolumn{1}{c}{} & \multicolumn{1}{c}{\multirow {4}{*}{NC}} & \multicolumn{1}{c}{$3/2^-(5/2^-)$} & \multicolumn{1}{c}{$23(15)$} & \multicolumn{1}{c}{} & \multicolumn{1}{c}{$23(15)$} & \multicolumn{1}{c}{} & \multicolumn{1}{c}{$23(15)$} \\
\cline{5-9}
\multicolumn{1}{l}{} & \multicolumn{1}{c}{} & \multicolumn{1}{c}{} & \multicolumn{1}{c}{$\nu^{}_e$} & \multicolumn{1}{c}{$3(1)$} & \multicolumn{1}{c}{} & \multicolumn{1}{c}{$4(3)$} & \multicolumn{1}{c}{} & \multicolumn{1}{c}{$4(2)$} \\
\multicolumn{1}{l}{} & \multicolumn{1}{c}{} & \multicolumn{1}{c}{} & \multicolumn{1}{c}{$\overline{\nu}^{}_e$} & \multicolumn{1}{c}{$3(2)$} & \multicolumn{1}{c}{} & \multicolumn{1}{c}{$4(2)$} & \multicolumn{1}{c}{} & \multicolumn{1}{c}{$4(3)$} \\
\multicolumn{1}{l}{} & \multicolumn{1}{c}{} & \multicolumn{1}{c}{} & \multicolumn{1}{c}{$\nu^{}_x$} & \multicolumn{1}{c}{$17(12)$} & \multicolumn{1}{c}{} & \multicolumn{1}{c}{$15(10)$} & \multicolumn{1}{c}{} & \multicolumn{1}{c}{$15(10)$} \\
\hline \hline
\end{tabular}
\vspace{0.2cm}
\caption{Neutrino events at JUNO for a core-collapse SN at a typical distance of 10 kpc. The gravitational binding energy of $3\times 10^{53}~{\rm erg}$ is assumed to be equally distributed in all six neutrino species, and the average neutrino energies are $\langle E^{}_{\nu_e}\rangle = 12~{\rm MeV}$, $\langle E^{}_{\overline{\nu}_e}\rangle = 14~{\rm MeV}$ and $\langle E^{}_{\nu_x}\rangle = 16~{\rm MeV}$. For $p$ES and $e$ES, a threshold of $0.2~{\rm MeV}$ for the recoil energy is chosen. {For ${^{13}}{\rm N}$-CC, the neutrino events for the ground state ${^{13}{\rm N}}(1/2^-)$ and the excited state ${^{13}{\rm N}}(3/2^-)$ are summed up. For ${^{13}}{\rm C}$-NC, the cross sections for the excited states ${^{13}}{\rm C}^*(3/2^-)$ and ${^{13}}{\rm C}^*(5/2^-)$ are dominant, and the neutrino events in the latter case are given in parentheses.}}
\label{table:events}
\end{table}
In a liquid-scintillator detector, there are various possibilities to probe SN neutrinos. For concreteness, we consider the JUNO detector~\cite{Li:2013zyd,An:2015jdp}. The detector is composed of 20 kiloton LAB-based liquid scintillator, for which the fraction of free protons is $12\%$. We employ the detector energy resolution of $3\%/\sqrt{E/{\rm MeV}}$, where $E$ is the visible energy of the prompt event signals (i.e., $E=E_{e^{+}}+m_{e}$). We also assume a detector threshold of $0.2~{\rm MeV}$ for the visible energies of recoiled protons and electrons. For a galactic SN at a distance of 10 kpc, the numbers of neutrino events in six primary channels at JUNO have been presented in Table~\ref{table:events}, where the SN neutrino fluences are given by Eq.~(1) under the assumption of the energy equipartition $E^{\rm tot}_{\nu_e} = E^{\rm tot}_{\overline{\nu}_e} = E^{\rm tot}_{\nu_x} = 5\times 10^{52}~{\rm erg}$. In addition, the average neutrino energies are $\langle E^{}_{\nu_e}\rangle = 12~{\rm MeV}$, $\langle E^{}_{\overline{\nu}_e}\rangle = 14~{\rm MeV}$ and $\langle E^{}_{\nu_x}\rangle = 16~{\rm MeV}$, while the spectral index $\gamma^{}_\alpha = 3$ is assumed to be universal for all neutrino flavors. Note that $\gamma^{}_\alpha = 2$ corresponds to the Maxwell-Boltzmann distribution, and numerical simulations indicate that the actual value $2 \lesssim \gamma^{}_\alpha \lesssim 4$ is time-dependent~\cite{Keil:2002in}. Hence, $\gamma^{}_\alpha = 3$ can be regarded as an effective description of the time-integrated energy spectra of SN neutrinos.
\begin{figure}
    \centering
    \includegraphics[width=0.8\textwidth]{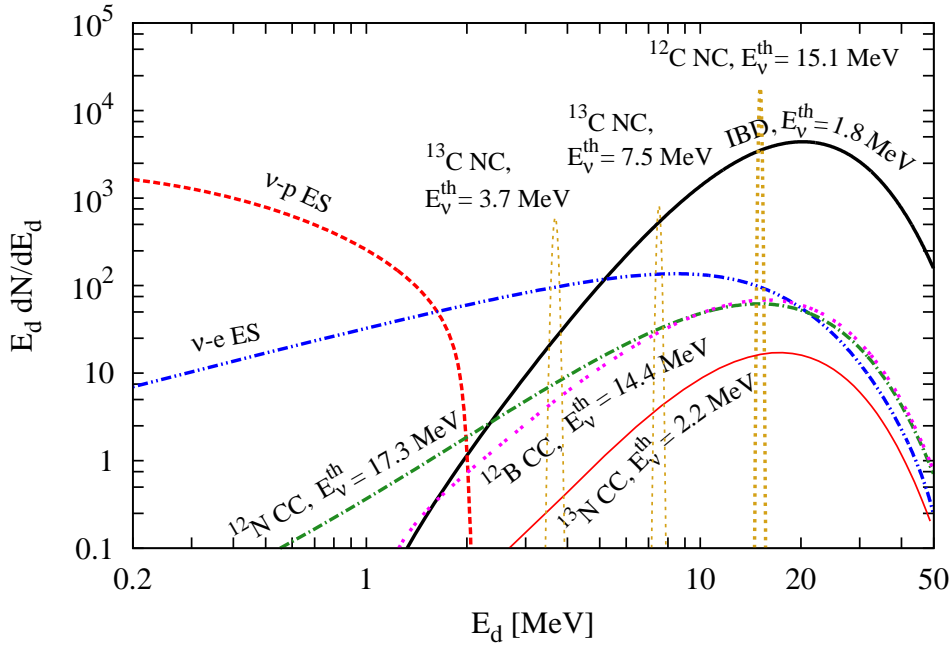}
    \vspace{-0.4cm}
\caption{The event spectra of prompt neutrino signals at JUNO for a galactic SN at a distance of 10 kpc, where $E^{}_{\rm d}$ stands for the visible energy in the detector. The SN neutrino fluxes are the same as in Table~\ref{table:events}.}
\end{figure}

The corresponding event spectra of prompt neutrino signals are shown in Fig. 1. In the following, the main features of the detection channels are summarized~\cite{An:2015jdp}:
\begin{enumerate}
\item The IBD is the most important channel for the detection of SN $\bar{\nu}_e$ in the {liquid-}scintillator detector, where a huge number of free protons are available. In this reaction, the neutrino energy threshold is $E^{\rm th}_\nu \approx \Delta + m_e \approx 1.806~{\rm MeV}$, where $\Delta \equiv m_n - m_p \approx 1.293~{\rm MeV}$ is the neutron-proton mass difference. The energy of the incident neutrino can be fully reconstructed from the positron energy via $E_\nu \approx E_{e^{+}} + \Delta$, as the recoil energies of nucleons are negligible. The annihilation of positrons and the capture of neutrons on free protons lead to a prompt and a delayed signal of gamma rays, respectively. Hence the time coincidence within $200~{\rm \mu s}$ of the prompt and delayed signals increases greatly the tagging power. In this work, we take the default selection efficiency of IBD events to be $95\%$, implying that $5\%$ of IBD events are detected without tagged neutrons. The latter could be an important background to the other channels, such as the $e$ES, as one can also observe from Fig.~1. To investigate the impact of IBD efficiency on the determination of total and average energies of $\nu^{}_e$ and $\nu^{}_x$, we also consider two cases of even lower and higher values, i.e., $90\%$ and $99\%$. Since the statistics of IBD events is already quite high, a slight reduction of the efficiency should not alter much the experimental sensitivities to $E^{\rm tot}_{\bar{\nu}_e}$ and $\langle E^{}_{\bar{\nu}_e} \rangle$.

    The IBD cross section has been precisely calculated in Ref.~\cite{Vogel:1999zy,Strumia:2003zx}, and applied to the detection of SN neutrinos. In general, the angular distribution of positrons is nearly isotropic, so it is difficult to extract the directional information of neutrinos. However, the angular distribution of neutrons may be used to further reduce backgrounds and locate the neutrino source~\cite{Vogel:1999zy,Apollonio:1999jg,Fischer:2015oma}.

\item The $p$ES channel is of crucial importance in detecting SN neutrinos of heavy-lepton flavors~\cite{Beacom:2002hs,Dasgupta:2011wg}. Although the total cross section of $p$ES is about four times smaller than that of IBD, all the neutrinos and antineutrinos of three flavors contribute and thus compensate for the reduction of cross section. In this channel, the recoil energy of proton $T_p \leq 2 E^2_\nu/m_p$ is highly suppressed by the nucleon mass, and will be further quenched in the scintillator matter. Therefore, the precise determination of quenching factor for proton and a low energy threshold are required to reconstruct neutrino energy and accumulate sufficient statistics. In our calculations, the Birks' law for the quenching effects of protons is implemented with a Birks' constant $0.0098 \pm 0.0003~{\rm cm}~{\rm MeV}^{-1}$, where the uncertainty is also included~\cite{vonKrosigk:2013sa}. The quench effects of positrons or electrons~\cite{Zhang:2014iza} are neglected in the current study.

    The $p$ES cross section was first calculated in Ref.~\cite{Weinberg:1972tu}, and has been recently simplified for low-energy neutrinos~\cite{Beacom:2002hs}. However, the cross section receives the dominant contribution from the axial form factor of proton, which at present is only known with a $30\%$ uncertainty if the strange-quark contribution to nucleon spin is taken into account~\cite{Ahrens:1986xe}. In the present work, an uncertainty of $20\%$ on the cross section has been considered for illustration. Low-energy neutrinos from pion decays at rest can be used to probe the proton strangeness and reduce the uncertainty of the axial charge by an order of magnitude in an underground laboratory with a kiloton {liquid-}scintillator detector~\cite{Pagliaroli:2012hq}.

\item In the $e$ES channel, the scattered electrons carry the directional information of incident neutrinos, and thus can be used to locate the SN. {This will be an extremely important approach in addition to the observation through the infrared light,} if a SN is hidden in the galactic gas and dust clouds and the optical signal is obscured. The $e$ES reaction is most sensitive to $\nu^{}_e$ because of its largest cross section, which is particularly useful in detecting the prompt $\nu^{}_e$ burst in the early stage of a SN explosion. However, it is difficult to determine the direction of a scattered electron in the {liquid-}scintillator detector after its multiple scattering. At this point, the large water-Cherenkov detectors, such as the SK and its upgraded version Hyper-Kamiokande, are necessary and complementary to the {liquid-}scintillator detectors. The cross sections of neutrino- and antineutrino-electron elastic scattering have been computed and summarized in Ref.~\cite{Marciano:2003eq}, where the electroweak radiative corrections are also included.

\item The neutral-current (NC) interaction on $^{12}{\rm C}$ is of crucial importance to probe neutrinos of heavy-lepton flavors, i.e., $\nu + {^{12}}{\rm C} \to \nu + {^{12}}{\rm C}^*$ (denoted as ${^{12}{\rm C}}$-NC), where $\nu$ collectively denotes neutrinos and antineutrinos of all three flavors. A $15.11$-MeV $\gamma$ from the deexcitation of $^{12}{\rm C}^*$ to its ground state is a clear signal of SN neutrinos. The cross section can be found in Refs.~\cite{Fukugita:1988hg,Volpe:2000zn}, and has also been measured in the LSND experiment~\cite{Auerbach:2001hz}. Since $\nu^{}_x$ has a higher average energy, the ${^{12}{\rm C}}$-NC channel is most sensitive to $\nu^{}_x$, offering a possibility to pin down the flavor content of SN neutrinos. However, the kinetic energy of $^{12}{\rm C}$ is heavily quenched in {liquid} scintillator, and thus invisible to current detectors, implying the impossibility to reconstruct neutrino energy event-by-event in this channel. In this sense, the $p$ES channel is more important. Note that the NC processes are not affected by the neutrino flavor oscillations.

\item As an advantage of the {liquid-}scintillator detector, the charged-current (CC) interaction on $^{12}{\rm C}$ takes place for both $\nu_e$ and $\overline{\nu}_e$ via
\begin{eqnarray}
&& \nu_e + {^{12}}{\rm C} \to e^- + {^{12}}{\rm N} \; , \label{eq: CCnue}\\
&& \overline{\nu}_e + {^{12}}{\rm C} \to e^+ + {^{12}}{\rm B} \; , \label{eq: CCnueb}
\end{eqnarray}
which will be denoted as ${^{12}{\rm N}}$-CC and ${^{12}{\rm B}}$-CC, respectively. The energy threshold for $\nu_e$ is $17.34~{\rm MeV}$, while that for $\overline{\nu}_e$ is $14.39~{\rm MeV}$. The subsequent beta decays of $^{12}{\rm B}$ and $^{12}{\rm N}$ with a $20.2~{\rm ms}$ and $11~{\rm ms}$ half-life, respectively, lead to a prompt-delayed coincident signal. Hence the CC reactions in Eqs.~(\ref{eq: CCnue}) and (\ref{eq: CCnueb}) provide another possibility to detect separately $\nu_e$ and $\overline{\nu}_e$.
However, the discrimination of electrons and positrons is difficult in large {liquid-}scintillator detectors, and we may only have the time and energy distributions to statistically separate the ${^{12}{\rm N}}$-CC and ${^{12}{\rm B}}$-CC processes. Therefore, we conservatively combine these two processes to one detection channel $^{12}{\rm C}$-CC in our current study. The cross section of neutrino interaction on $^{12}{\rm C}$ has been calculated in Ref.~\cite{Fukugita:1988hg} by using a direct evaluation of nuclear matrix elements from experimental data at that time. Recent calculations based on the nuclear shell model and the random-phase approximation can be found in Ref.~\cite{Volpe:2000zn}. The cross section has been measured in the LSND experiment, and the result is well compatible with theoretical calculations~\cite{Auerbach:2001hz}.

\item The ${^{13}}{\rm C}$ atoms in the liquid scintillator can also be implemented to capture SN neutrinos, although the natural abundance of ${^{13}}{\rm C}$ is about $1.1\%$. Similar to the case of ${^{12}}{\rm C}$, both CC and NC interactions of neutrinos and antineutrinos should be considered. The numbers of neutrino events are in general small in the ${^{13}}{\rm C}$ case, so we include the corresponding reactions in the determination of $\nu^{}_e$ and $\nu^{}_x$ properties.

    For the CC interaction, we consider
    \begin{eqnarray}
    \nu^{}_e + {^{13}}{\rm C} \to e^- + {^{13}}{\rm N} \; ,
    \end{eqnarray}
    where the final state ${^{13}}{\rm N}$ can be either the ground state $1/2^-$ or the excited state $3/2^-$. The energy threshold in the ground-state case is $E^{\rm th}_\nu \approx 2.2~{\rm MeV}$, while the excitation energy of ${^{13}}{\rm N}^*(3/2^-)$ is about 3.51 MeV. The cross sections for higher excited states are suppressed~\cite{Fukugita:1989wv,Suzuki:2012aa}, so those reactions can be safely ignored. As the de-excitation of ${^{13}}{\rm N}^*(3/2^-)$ to the ground state is extremely rapid and the beta-decay lifetime of ${^{13}}{\rm N}$ is about ten minutes, it is impossible to distinguish between ${^{13}}{\rm N}$-CC and $e$ES signals. Therefore, we combine these two channels together and denote as $e$ES+${^{13}}{\rm N}$-CC.

    For the NC reaction
    \begin{eqnarray}
    \nu + {^{13}}{\rm C} \to \nu + {^{13}}{\rm C}^* \; ,
    \end{eqnarray}
    we take account of two excited states $3/2^-$ and $5/2^-$ of ${^{13}}{\rm C}$, whose excitation energies are 3.685 MeV and 7.547 MeV, respectively. For SN neutrinos, the cross sections for these two states are dominant over those for higher excited states~\cite{Fukugita:1989wv,Suzuki:2012aa}. The signals for this reaction ${^{13}}{\rm C}$-NC are the same as those for ${^{12}}{\rm C}$-NC, but with the de-excitation photons of different energies.
\end{enumerate}

\subsection{Flavor Conversions}

The flavor conversions of SN neutrinos are complicated by the dense matter and neutrino background. In the SN core, the matter density is so high that lepton flavors are approximately conserved due to the frequent scattering of neutrinos with background particles~\cite{Hannestad:1999zy}. From the neutrino-sphere up to one thousand kilometers, the neutrino-neutrino refraction may lead to spectral splits of SN neutrinos~\cite{Duan:2010bg}. However, it remains to be clarified whether the nonlinear collective neutrino oscillations do take place in a real supernova environment. For simplicity, we temporarily put aside collective neutrino oscillations, but keep in mind that they may have important impact on the SN neutrino detection~\cite{Mirizzi:2015eza}. Even farther out from the SN core, the Mikheyev-Smirnov-Wolfenstein (MSW) matter effects~\cite{Mikheev:1986gs,Mikheev:1986wj,Wolfenstein:1977ue} come into play and reprocess neutrino spectra~\cite{Dighe:1999bi}. According to the latest neutrino oscillation data~\cite{Agashe:2014kda}, two neutrino mass-squared differences $\Delta m^2_{21} \equiv m^2_2 - m^2_1 \approx 7.5\times 10^{-5}~{\rm eV}^2$ and $|\Delta m^2_{31}| \equiv |m^2_3 - m^2_1| \approx 2.4\times 10^{-3}~{\rm eV}^2$ are precisely measured, and the reactor neutrino mixing angle $\theta^{}_{13}$ is found to be relatively large (i.e., $\sin^2 \theta^{}_{13} \approx 0.024$). The resonant flavor conversions corresponding to two neutrino mass-squared differences in the outer layer of a SN are perfectly adiabatic for the observed mixing angles, i.e., $\sin^2 \theta^{}_{12} \approx 0.303$ and $\sin^2 \theta^{}_{13} \approx 0.024$, as shown in Ref.~\cite{Dighe:1999bi}. After flavor conversions, the fluxes of neutrinos leaving the SN are related to the initial ones as
\begin{eqnarray}
F^{}_{\nu^{}_e} &=& F^0_{\nu^{}_x} \; , \nonumber \\
F^{}_{\overline{\nu}^{}_e} &=& \cos^2 \theta^{}_{12} F^0_{\rm \overline{\nu}^{}_e} + \sin^2 \theta^{}_{12} F^0_{\nu^{}_x} \; , \nonumber \\
F^{}_{\nu^{}_x} &=& \frac{1}{4} \left(2 + \cos^2 \theta^{}_{12}\right) F^0_{\nu^{}_x} + \frac{1}{4} F^0_{\nu^{}_e} + \frac{1}{4} \sin^2 \theta^{}_{12} F^0_{\overline{\nu}^{}_e} \; ,
\end{eqnarray}
for the normal neutrino mass ordering with $\Delta m^2_{31} > 0$ (NO); and
\begin{eqnarray}
F^{}_{\nu^{}_e} &=& \sin^2 \theta^{}_{12} F^0_{\nu^{}_e} + \cos^2 \theta^{}_{12} F^0_{\nu^{}_x} \; , \nonumber \\
F^{}_{\overline{\nu}^{}_e} &=&  F^0_{\nu^{}_x} \; , \nonumber \\
F^{}_{\nu^{}_x} &=& \frac{1}{4} \left(2 + \sin^2 \theta^{}_{12}\right) F^0_{\nu^{}_x} + \frac{1}{4} \cos^2 \theta^{}_{12} F^0_{\nu^{}_e} + \frac{1}{4} F^0_{\overline{\nu}^{}_e} \; ,
\end{eqnarray}
for the inverted neutrino mass ordering with $\Delta m^2_{31} < 0$ (IO). Note that the fluxes $F^{}_\alpha$ at a distance $D$ should be scaled by a factor of $r^2/D^2$, where the initial fluxes $F^0_\alpha$ are evaluated at a radius $r$ to the SN core.

In order to investigate the experimental sensitivity to the average energy and the total energy of each flavor, we first assume that there are no flavor oscillations in Sec. III. In this case, the neutrino spectrum for a given flavor is parameterized by the average energy $\langle E^{}_\alpha \rangle$ and the total energy $E^{\rm tot}_\alpha$. In the presence of flavor conversions, the neutrino spectra at the Earth are no longer thermal, as we have seen in Eqs.~(6) and (7). Therefore, we turn to the realistic case where neutrino flavor conversions are included, and test the hypothesis of energy equipartition at the future large {liquid}-scintillator detectors in Sec. IV. {We also provide the Appendix for the details of the $\chi^2$ functions and the corresponding efficiencies, backgrounds, and systematics in our calculations.}

\section{Different Neutrino Flavors}

\subsection{The $\overline{\nu}^{}_e$ Spectrum}

\begin{figure}
  \begin{minipage}[t]{0.5\textwidth}
    \vspace{0pt}
    \centering
    \includegraphics[width=\textwidth]{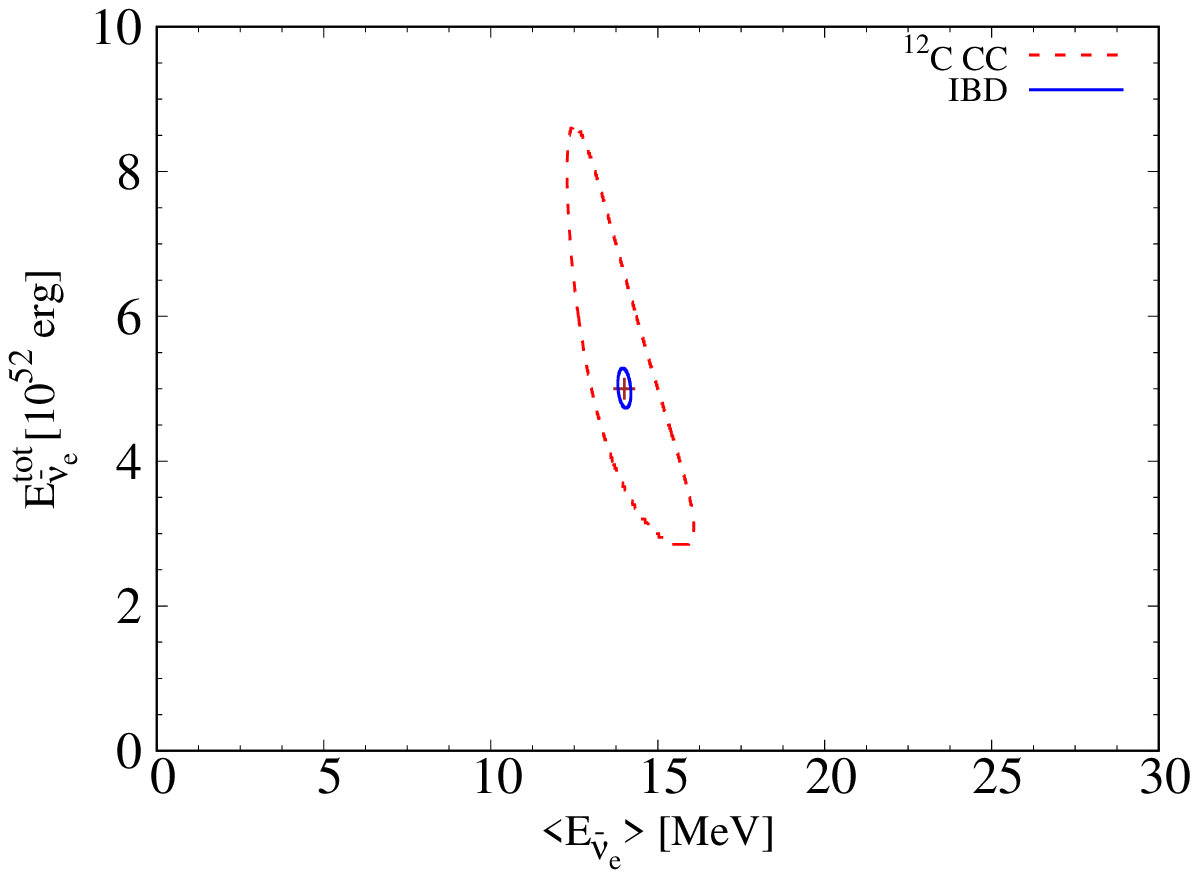}
  \end{minipage}%
  \begin{minipage}[t]{0.5\textwidth}
    \vspace{0pt}
    \centering
    \includegraphics[width=\textwidth]{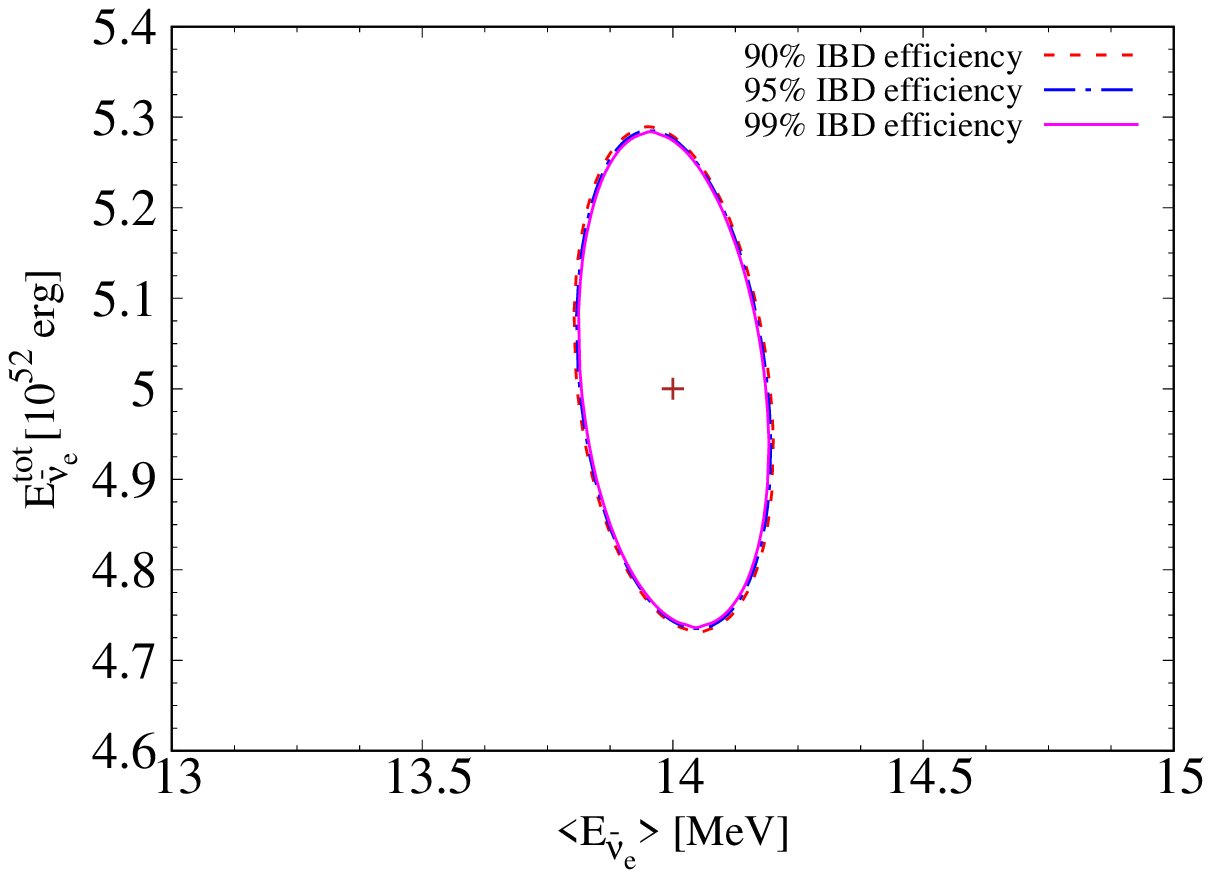}
  \end{minipage}
\caption{Allowed regions of the luminosity ($E^{\rm tot}_{\overline{\nu}_e}$) and average energy ($\langle E^{}_{\overline{\nu}_e}\rangle$) of $\overline{\nu}^{}_e$
at the $90\%$ C.L. in the individual (left panel) and global (right panel) fitting of the IBD and $^{12}{\rm C}$-CC processes. The left panel uses an IBD efficiency of 95\%, whereas a variation of the IBD efficiency has been applied in the right panel. Note that {the crosses stand for the best-fit values, and }the right panel is rescaled for better visibility.}
\label{fig:nuebar}
\end{figure}
The $\overline{\nu}^{}_e$ flavor is detected in the IBD and ${^{12}}{\rm C}$-CC processes, which according to Table~\ref{table:events} have the event statistics of around 5000 and 100, respectively. Therefore, the parameters of the $\overline{\nu}^{}_e$ flavor are determined predominately by the IBD channel, which is shown in the left panel of Fig.~\ref{fig:nuebar}. The precision of the luminosity (i.e., $E^{\rm tot}_{\overline{\nu}_e}$) and average energy (i.e., $\langle E^{}_{\overline{\nu}_e}\rangle$) at the $90\%$ confidence level (C.L.) in the IBD channel are $5.4\%$ and $1.4\%$, respectively. In the right panel of Fig.~\ref{fig:nuebar}, we illustrate the global fitting regions of the IBD and $^{12}{\rm C}$-CC processes at the $90\%$ C.L. with different IBD selection efficiencies, which are proved to have negligible effects on the $E^{\rm tot}_{\overline{\nu}_e}$ and $\langle E^{}_{\overline{\nu}_e}\rangle$ determination. This is mainly because the statistical uncertainty is much smaller than the corresponding systematic uncertainty in the IBD detection channel.

\subsection{The $\nu^{}_e$ Spectrum}

\begin{figure}
  \begin{minipage}[t]{0.5\textwidth}
    \vspace{0pt}
    \centering
    \includegraphics[width=\textwidth]{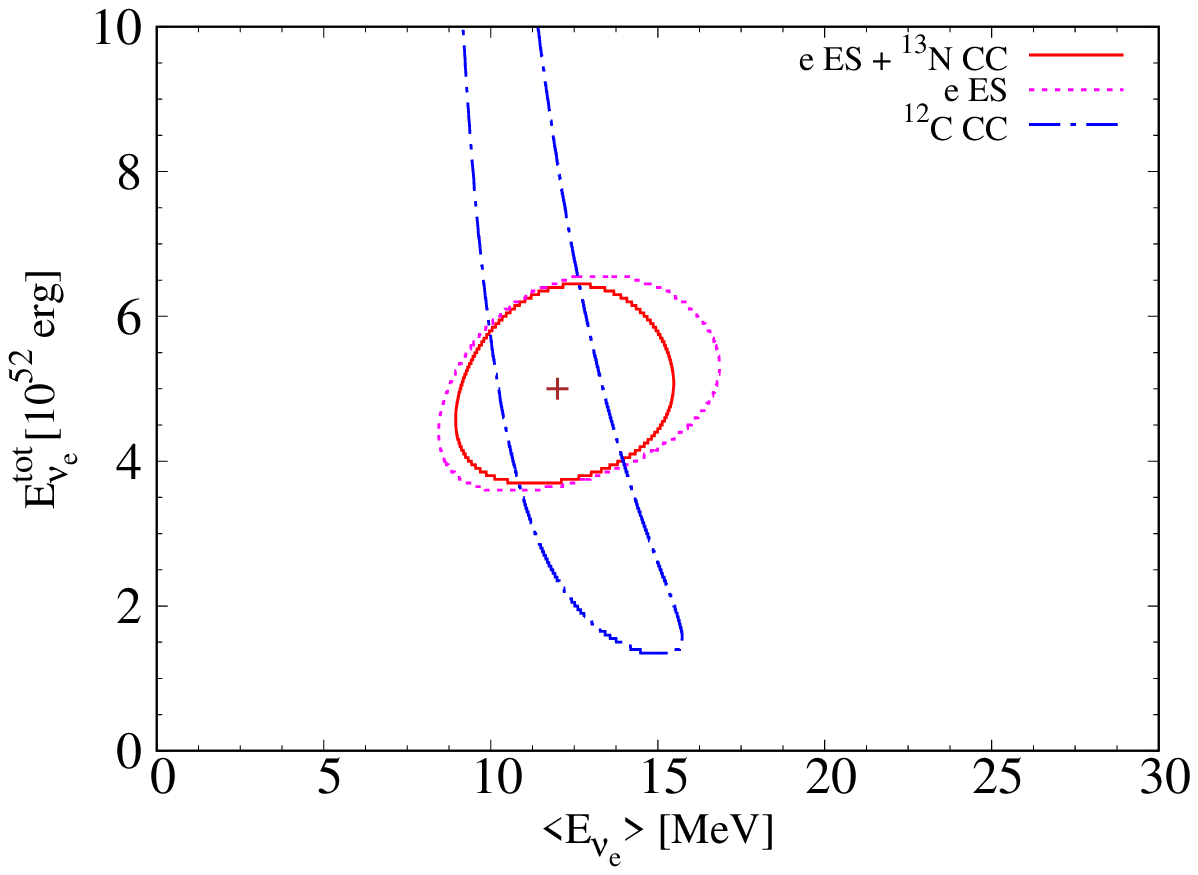}
  \end{minipage}%
  \begin{minipage}[t]{0.5\textwidth}
    \vspace{0pt}
    \centering
    \includegraphics[width=\textwidth]{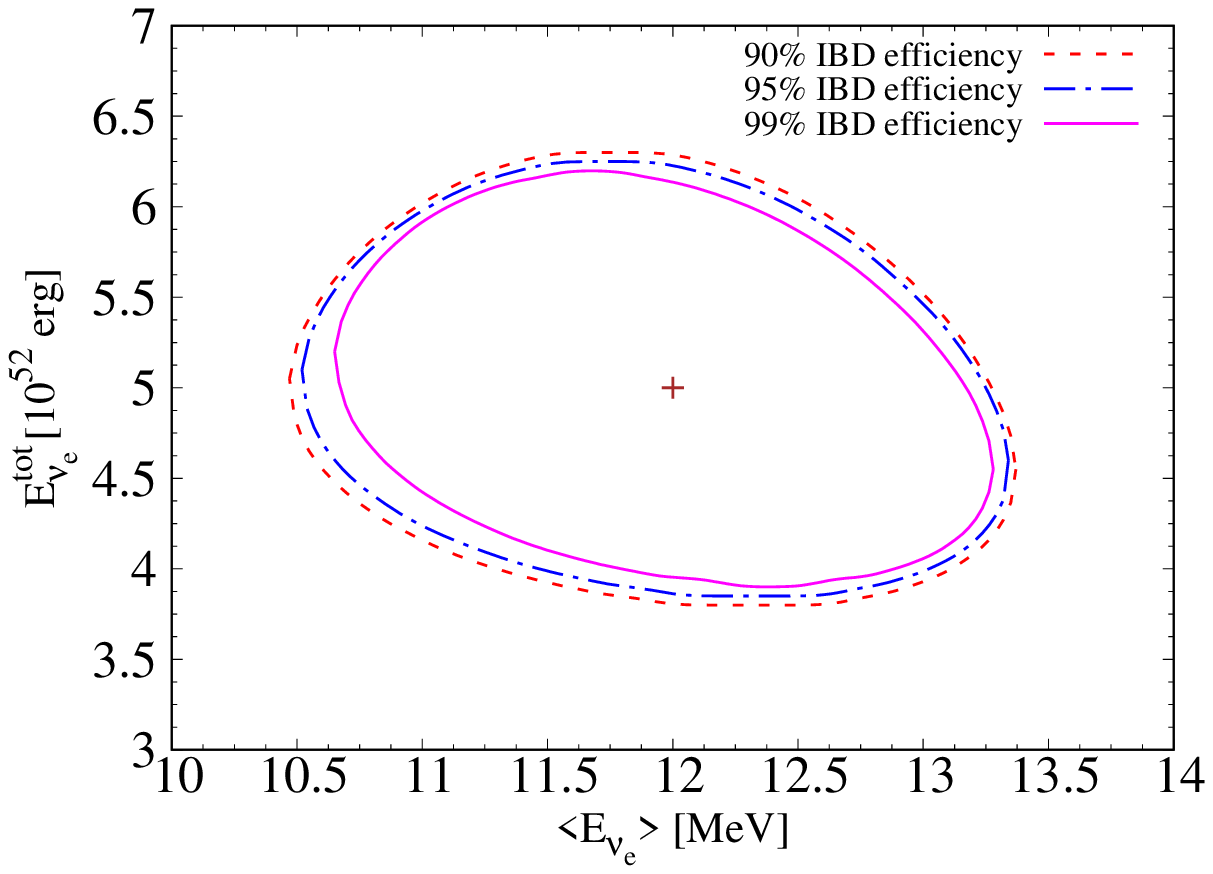}
  \end{minipage}
\caption{Allowed regions of the luminosity ($E^{\rm tot}_{{\nu}_e}$) and average energy ($\langle E^{}_{{\nu}_e}\rangle$) of ${\nu}^{}_e$
at the $90\%$ C.L. in the individual (left panel) and global (right panel) fitting of the $e$ES+{$^{13}{\rm N}$-CC} and $^{12}{\rm C}$-CC processes. The left panel uses an IBD efficiency of $95\%$, whereas variations of the IBD efficiency have been applied in the right panel. Note that {the crosses stand for the best-fit values, and} the right panel is rescaled for better visibility.}
\label{fig:nue}
\end{figure}

The ${\nu}^{}_e$ flavor contributes to the $e$ES, $^{12}{\rm C}$-CC, $^{12}{\rm C}$-NC, {$^{13}{\rm N}$-CC and $^{13}{\rm C}$-NC} processes, and is correspondingly constrained by the measurements in these channels. In Fig.~\ref{fig:nue}, we show the allowed regions of the luminosity ($E^{\rm tot}_{{\nu}_e}$) and average energy ($\langle E^{}_{{\nu}_e}\rangle$) of ${\nu}^{}_e$ at the $90\%$ C.L. in the individual (left panel) and global (right panel) fitting of the $e$ES, $^{12}{\rm C}$-CC and {$^{13}{\rm N}$-CC} processes. {To illustrate the impact of $^{13}{\rm N}$-CC, we present two separate contours for the case with only $e$ES (pink and dashed curve) and the combined case of $e$ES+$^{13}{\rm N}$-CC (red and solid curve). One can observe that the inclusion of $^{13}{\rm N}$-CC events indeed improves the results.} The left panel uses an IBD efficiency of $95\%$, whereas variations of the IBD efficiency have been applied in the right panel. In the left panel, we can observe that both the $e$ES and $^{12}{\rm C}$-CC processes are sensitive to the ${\nu}^{}_e$ energy spectrum, providing an excellent measurement of the ${\nu}^{}_e$ average energy. On the other hand, because the event number of the $e$ES {(plus $^{13}{\rm N}$-CC)} process is much larger than that of the $^{12}{\rm C}$-CC process, the $e$ES channel has a better sensitivity to the ${\nu}^{}_e$ luminosity, which turns out to be $27\%$ at the $90\%$ C.L. The $^{12}{\rm C}$-NC {and $^{13}{\rm C}$-NC} process only measure a total rate of the neutrino fluxes, {so we expect that the allowed region displays an anti-correlation relation between the luminosity and average energy of ${\nu}^{}_e$. However, the main contributions to these two kinds of signals actually come from $\nu^{}_x$, as shown in Table~\ref{table:events}. Therefore, they set very poor constraints on the average energy and luminosity of $\nu^{}_e$.} Combining {three dominant channels}, we obtain the global fitting results in the right panel of Fig.~\ref{fig:nue}. The accuracies of $E^{\rm tot}_{{\nu}_e}$ and $\langle E^{}_{{\nu}_e}\rangle$ at the $90\%$ C.L. are {$24\%$ and $12\%$}, respectively. With the decreasing IBD efficiencies, backgrounds from the increasing un-tagged positrons of IBD samples would have a sizeable effect on the measurements of the luminosity and average energy of ${\nu}^{}_e$.

\subsection{The $\nu^{}_x$ Spectrum}

\begin{figure}
  \begin{minipage}[t]{0.5\textwidth}
    \vspace{0pt}
    \centering
    \includegraphics[width=\textwidth]{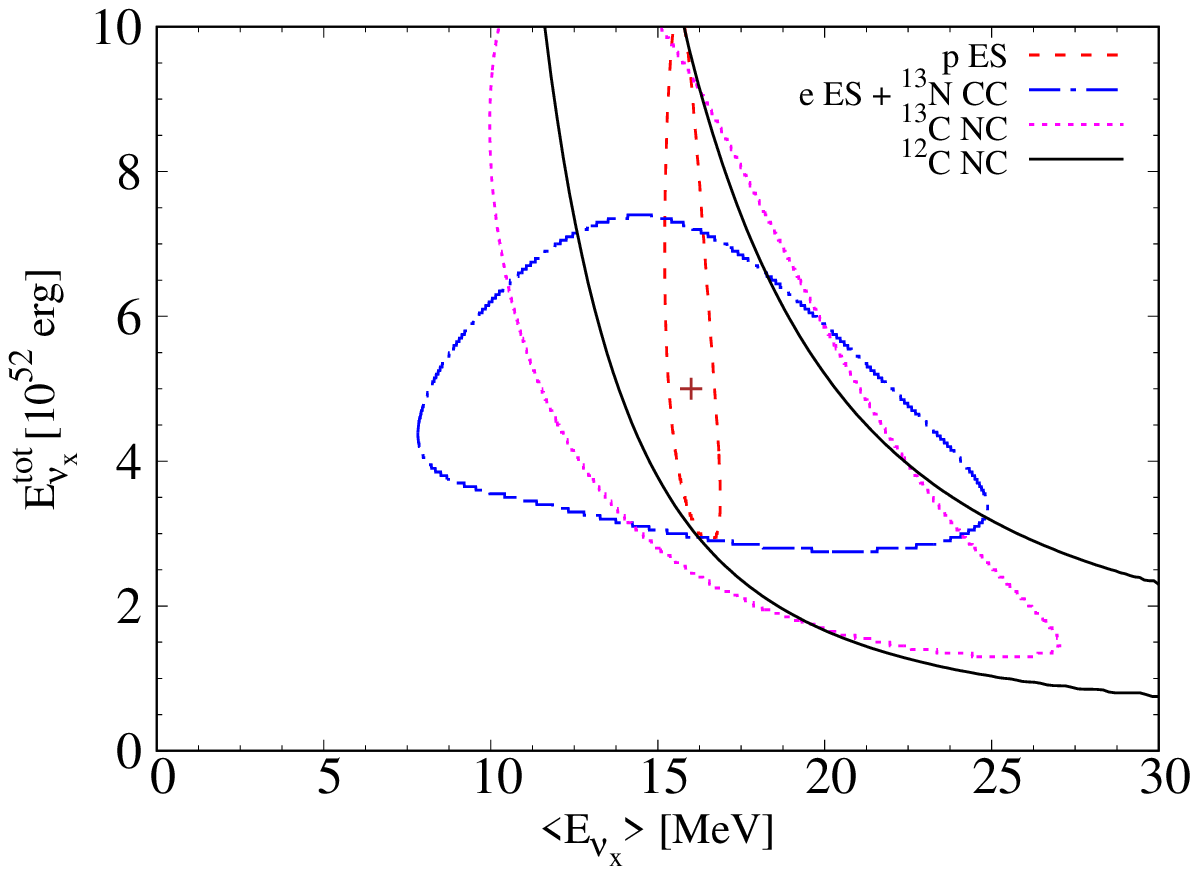}
  \end{minipage}%
  \begin{minipage}[t]{0.5\textwidth}
    \vspace{0pt}
    \centering
    \includegraphics[width=\textwidth]{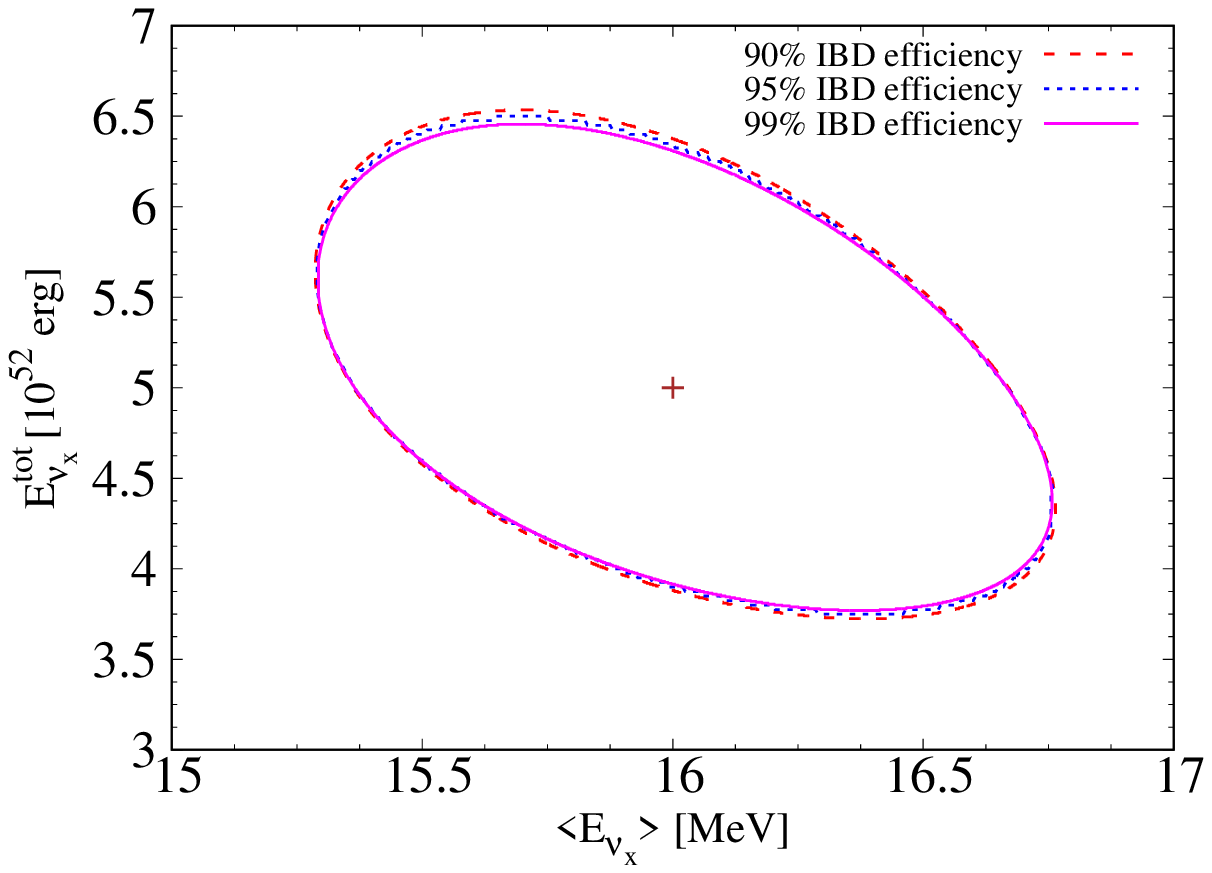}
  \end{minipage}
\caption{Allowed regions of the luminosity ($E^{\rm tot}_{{\nu}_x}$) and average energy ($\langle E^{}_{{\nu}_x}\rangle$) of ${\nu}^{}_x$
at the $90\%$ C.L. in the individual (left panel) and global (right panel) fitting of the $e$ES{+$^{13}{\rm N}$-CC}, $p$ES, $^{12}{\rm C}$-NC {and $^{13}{\rm C}$-NC} processes. The left panel uses an IBD efficiency of $95\%$, whereas variations of the IBD efficiency have been applied in the right panel. Note that {the crosses stand for the best-fit values, and }the right panel is rescaled for better visibility.}
\label{fig:nux}
\end{figure}
The ${\nu}^{}_x$ flavor is measured in the NC interaction channels, namely, the $e$ES+{$^{13}{\rm N}$-CC}, $p$ES, $^{12}{\rm C}$-NC {and $^{13}{\rm C}$-NC}. In Fig.~\ref{fig:nux}, we illustrate the allowed regions of the luminosity ($E^{\rm tot}_{{\nu}_x}$) and average energy ($\langle E^{}_{{\nu}_x}\rangle$) at the 90\% C.L. in the individual (left panel) and global (right panel) fitting of the $e$ES{+$^{13}{\rm N}$-CC}, $p$ES, $^{12}{\rm C}$-NC {and $^{13}{\rm C}$-NC} processes. The left panel uses an IBD efficiency of $95\%$, whereas variations of the IBD efficiency have been applied in the right panel. The $p$ES channel in Table~\ref{table:events} contains around 1500 events above the 0.2 MeV energy threshold, which offers the most precision measurement of the ${\nu}^{}_x$ average energy. The precision level of $\langle E^{}_{{\nu}_x}\rangle$ at the $90\%$ C.L. is {$5.2\%$}. As for the luminosity, however, it is shown that the $p$ES is not the best probe because a $20\%$ uncertainty of the cross section normalization has been included. On the other hand, the $e$ES process, which has the smallest normalization uncertainty, produces the most accurate measurement of the luminosity. The $p$ES gives a comparable precision for the lower limit of $E^{\rm tot}_{{\nu}_x}$, but the upper limit is much worse. {The $^{13}{\rm C}$-NC channel offers a slightly better constraint, compared to $^{12}{\rm C}$-NC, since the former contains the spectrum information from two distinct peaks.} Combining these complementary processes, we obtain the global fitting results in the right panel of Fig.~\ref{fig:nux}, where the precision levels of the luminosity and average energy at the $90\%$ C.L. turn out to be {$27\%$ and $4.6\%$}, respectively. We also observe that variations of the IBD efficiency have negligible impact on the precision of $\langle E^{}_{{\nu}_x}\rangle$ but small visible effects on $E^{\rm tot}_{{\nu}_x}$. This is because $\langle E^{}_{{\nu}_x}\rangle$ is predominately determined from the measurement of the $p$ES process and $E^{\rm tot}_{{\nu}_x}$ is constrained from all the three channels.

{Before ending this section, we make some comments on the adopted energy threshold $0.2~{\rm MeV}$ for the single events, and on the impact of the sizes of liquid-scintillator detectors. First, although such a low threshold is achieved in Borexino, the currently largest liquid-scintillator detector at KamLAND reaches only $0.7~{\rm MeV}$. For KamLAND, the primary goal is to detect reactor antineutrinos, for which the threshold for the prompt energy of the IBD events in the detector is around $1.0~{\rm MeV}$, so a cutoff at $0.7~{\rm MeV}$ is good enough to cover all the antineutrino events. If this cutoff is also applied to JUNO, the number of $p$ES events will be reduced from 1500 to 250, implying that the precisions of $\nu^{}_x$ average energy and luminosity will be worsen by a factor of $\sqrt{6} \approx 2.4$ if only the statistical error is considered.
If an intermediate energy threshold at $0.5~{\rm MeV}$ is achieved, the $p$ES event number will be around 400, and the parameter precision will be reduced by a factor of $\sqrt{3.5} \approx 1.9$. Second, since the target mass of the proposed RENO-50 experiment is 18 kiloton, our results for JUNO are also applicable to RENO-50. For the LENA detector, the target mass will be larger by a factor of 2.5, so we expect a precision enhanced by a factor of $\sqrt{2.5} \approx 1.6$ for the channel where the statistical error dominates. This is the case for the SN $\nu^{}_e$ and $\nu^{}_x$, but the systematic uncertainty may be important for $\overline{\nu}^{}_e$ due to a huge number of IBD events.}

\section{Hypothesis of Energy Equipartition}

After exploring the experimental capability to determine the average energy and luminosity of SN neutrinos for each flavor, we proceed to consider the potential of future large {liquid-}scintillator detectors to test fundamental assumptions in SN physics. One of the most important assumptions is that the gravitational binding energy of $3\times 10^{53}~{\rm erg}$ is equally distributed in all neutrino species (i.e., $\nu^{}_e$, $\nu^{}_\mu$, $\nu^{}_\tau$ and their antiparticles), which is known as the hypothesis of energy equipartition and has been taken for granted in most works related to SN neutrinos. Since $\nu^{}_\mu$ and $\overline{\nu}^{}_\mu$ interact almost in the same way with matter and they are produced in pairs via neutral-current interactions, and likewise for $\nu^{}_\tau$ and $\overline{\nu}^{}_\tau$, it should be reasonable to assume that their energies are equal. However, for $\nu^{}_e$ and $\overline{\nu}^{}_e$, which possess charged-current interactions and are produced in different ways, it is not well justified that they should have the same energy as the neutrinos of heavy flavors.
\begin{figure}
\begin{center}
\begin{tabular}{c}
\includegraphics[width=0.6\textwidth,angle=-90]{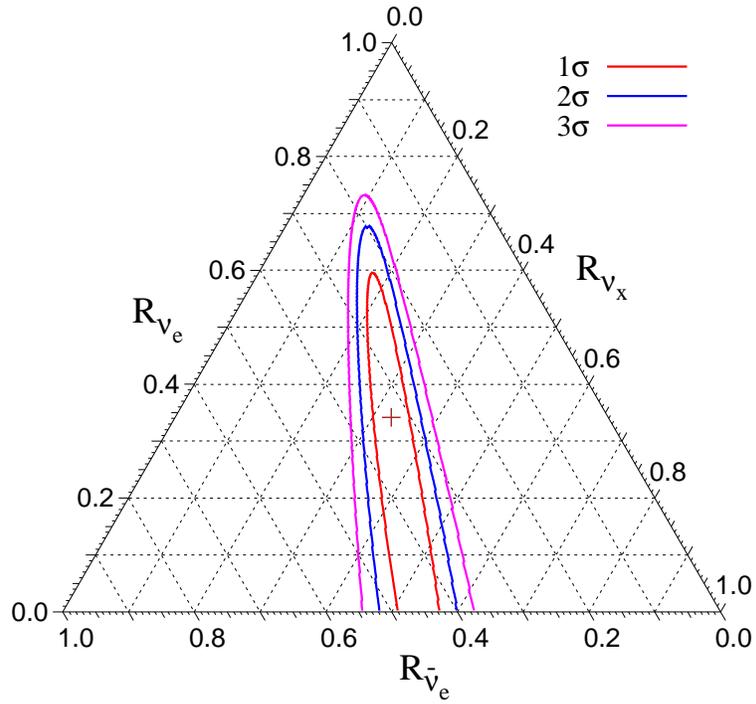}
\\
\includegraphics[width=0.6\textwidth,angle=-90]{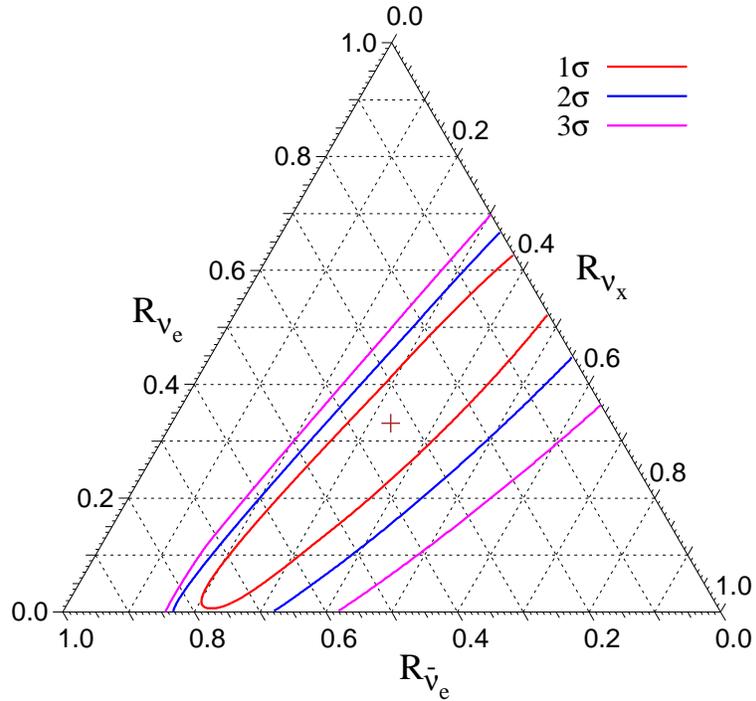}
\end{tabular}
\end{center}
\vspace{-0.6cm}
\caption{Allowed regions of the neutrino energy ratios ($R^{}_{\nu^{}_{e}}$, $R^{}_{\overline{\nu}_e}$, $R^{}_{\nu^{}_{x}}$) at the 1$\sigma$, 2$\sigma$ and 3$\sigma$ C.L. are shown for the NO (upper panel) and IO (lower panel) cases, respectively. The crosses stand for the best fit values.
\label{fig:ternary}}
\end{figure}

In order to test the energy-equipartition hypothesis, we allow $E^{\rm tot}_{{\nu}_e}$, $E^{\rm tot}_{\overline{\nu}_e}$ and $E^{\rm tot}_{{\nu}_x}$ to be different, and define the total energy of all flavor neutrinos as $E^{\rm tot}_{} = E^{\rm tot}_{{\nu}_e} + E^{\rm tot}_{\overline{\nu}_e} + 4E^{\rm tot}_{{\nu}_x}$, and the energy ratio of the flavor neutrino $\nu^{}_{\alpha}$ as
\begin{eqnarray}
R^{}_{\nu^{}_{\alpha}} = \frac{E^{\rm tot}_{\nu^{}_\alpha}}{E^{\rm tot}_{\nu^{}_e}+E^{\rm tot}_{\overline{\nu}_e}+E^{\rm tot}_{\nu^{}_x}}\;.
\end{eqnarray}
In this section, we take the same assumptions as in previous numerical calculations with $E^{\rm tot}_{}=3\times 10^{53}~{\rm erg}$ and ($R^{}_{\nu^{}_{e}}$, $R^{}_{\overline{\nu}_e}$, $R^{}_{\nu^{}_{x}}$)=(1/3, 1/3, 1/3). Different from Sec.~III, we employ the realistic neutrino flavor conversions as shown in Eqs.~(6) and (7) for this study. In testing the energy-equipartition hypothesis, {assuming the $^{12}{\rm N}$-CC and $^{12}{\rm B}$-CC processes are indistinguishable, and $e$ES and $^{13}{\rm N}$-CC are combined to $e$ES+$^{13}{\rm N}$-CC,} we consider the global fitting of all the detection channels with proper correlation of the detection systematic uncertainties. The statistical fitting results for the energy ratios are presented in Fig.~5 in the form of ternary plots, where the allowed regions of $R^{}_{\nu^{}_{e}}$, $R^{}_{\overline{\nu}_e}$ and $R^{}_{\nu^{}_{x}}$ at the 1$\sigma$, 2$\sigma$ and 3$\sigma$ C.L. are shown for the NO (upper panel) and IO (lower panel) cases, respectively. {Note that the value of $R^{}_{\nu^{}_\alpha}$ should be read off according to the ticks along the corresponding side of the triangle.} The 1$\sigma$ ranges for $R^{}_{\nu^{}_{e}}$, $R^{}_{\overline{\nu}_e}$ and $R^{}_{\nu^{}_{x}}$ are respectively {[0, 0.60], [0.23, 0.49] and [0.17, 0.57] in the NO case, while [0.01, 0.63], [0, 0.78] and [0.20, 0.48] in the IO case.} Comparing the upper and lower panels, the orientation of allowed regions are totally different for two cases. The NO case has the best measurement of $R^{}_{\overline{\nu}_e}$, whereas the most accurate measurement of $R^{}_{\nu^{}_{x}}$ is obtained in the IO case. This property can be understood with the help of Eqs.~(4) and (5), where one can notice that $70\%$ ($100\%$) of the initial ${\overline{\nu}_e}$ (${{\nu}_x}$) fluxes are measured in the IBD process. On the other hand, we can only obtain the upper bound for $R^{}_{\nu^{}_{e}}$ in the NO case and for $R^{}_{\overline{\nu}_e}$ in the IO case. Because of the neutrino flavor conversions, the initial $\nu^{}_{e}$ flavor in the NO case and the initial $\overline{\nu}_e$ flavor in the IO case are both recognized as $\nu^{}_{x}$ in the NC interaction processes. Therefore, their contributions can be compensated by the fluctuation of other initial neutrino flavors (i.e., $\overline{\nu}_e$ and $\nu^{}_{x}$ in the NO case and ${\nu}_e$ and $\nu^{}_{x}$ in the IO case).

\begin{figure}
\begin{center}
\begin{tabular}{c}
\includegraphics[width=0.65\textwidth]{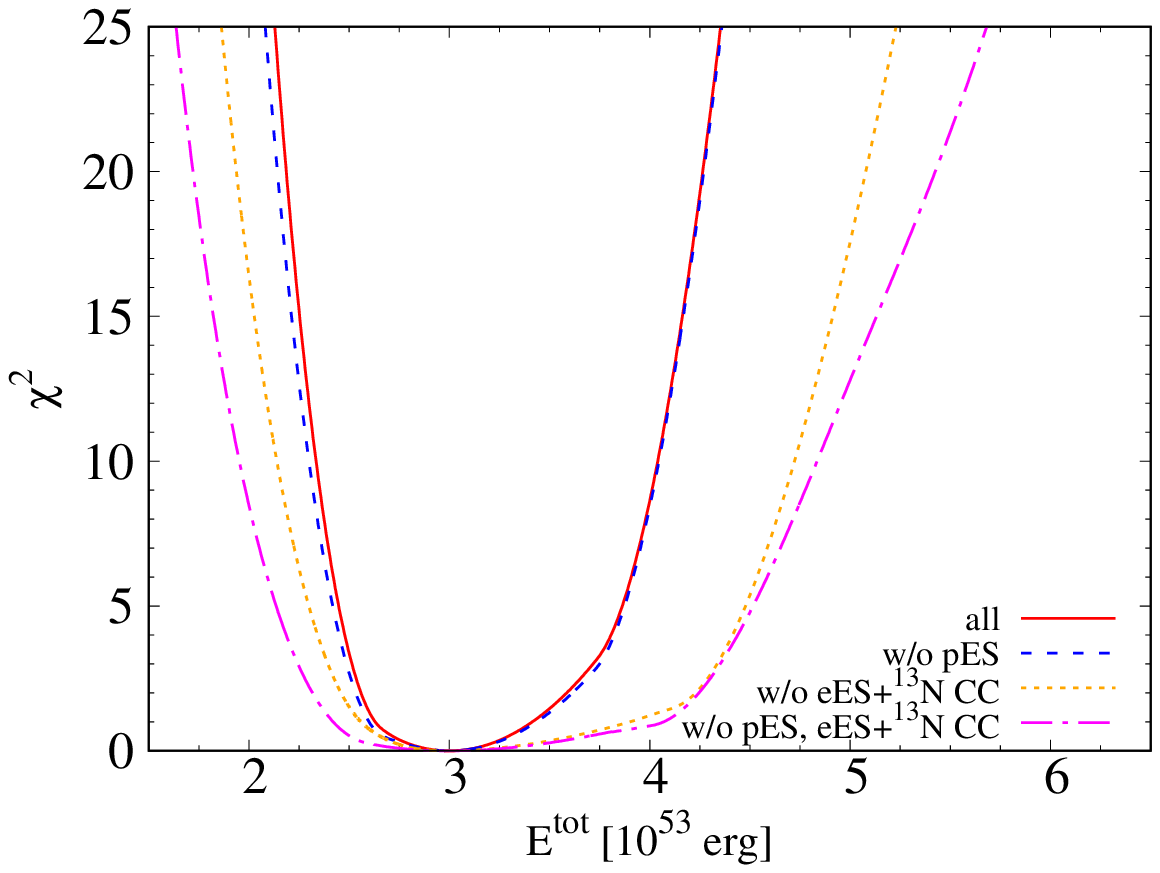}
\\
\includegraphics[width=0.65\textwidth]{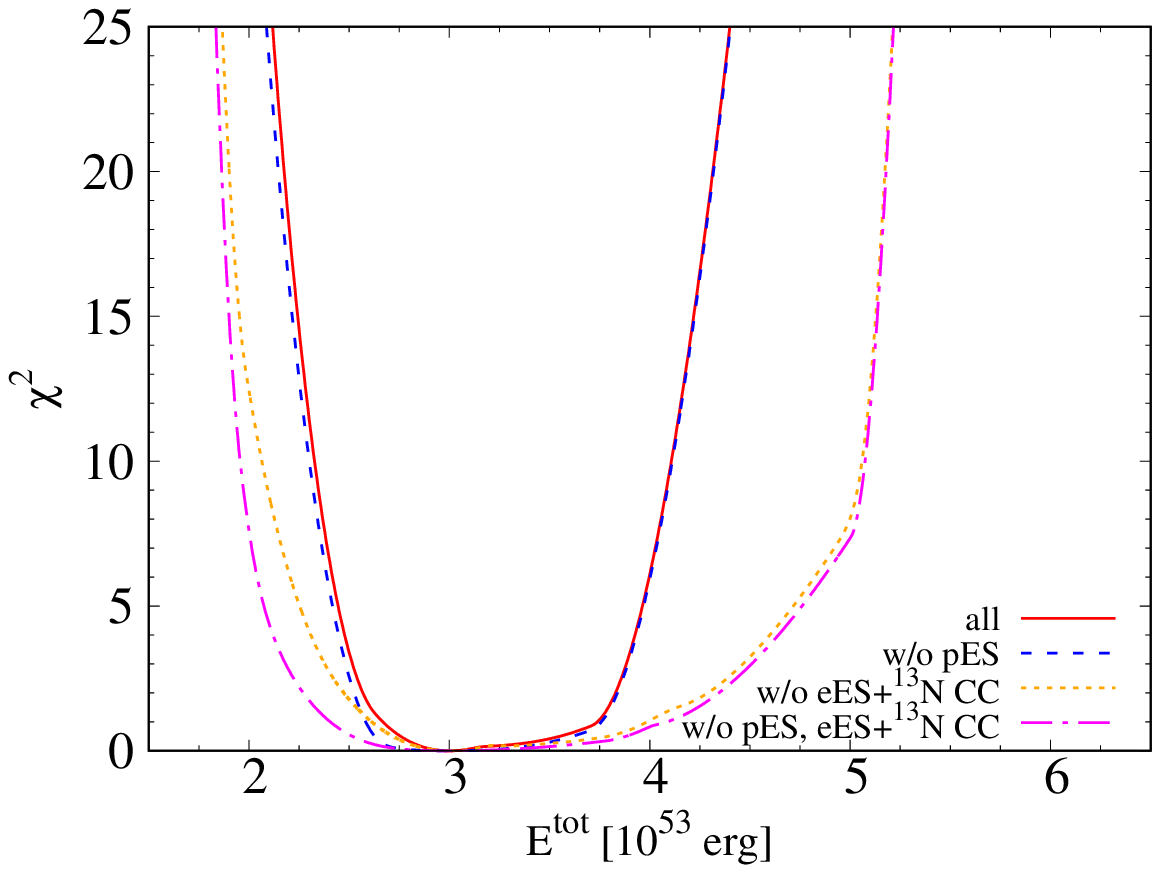}
\end{tabular}
\end{center}
\vspace{-0.6cm} \caption{The $\Delta \chi^2$ distributions for the gravitational binding energy $E^{\rm tot}_{}$ in the
NO (upper panel) and IO (lower panel) cases in the global-fitting scenarios. The solid lines, short dashed lines, dotted lines,
and dash-dotted lines stand for the cases of using all detection processes, without $p$ES process,
without the $e$ES+{$^{13}{\rm N}$-CC} processes, and without both $p$ES and $e$ES+{$^{13}{\rm N}$-CC} processes, respectively.
\label{fig:lumi}}
\end{figure}
Besides testing of the energy-equipartition hypothesis, it is also interesting to constrain the total energy of all flavor neutrinos $E^{\rm tot}_{}$ in the global-fitting scenarios. In Fig.~6, we illustrate the $\Delta \chi^2$ distributions of $E^{\rm tot}_{}$ in the NO (upper panel) and IO (lower panel) cases. The solid lines, short dashed lines, dotted lines, and dash-dotted lines are for the cases of using all detection processes, without the $p$ES process, without the $e$ES process, and without both $e$ES and $p$ES processes, respectively. From the figure, we observe sudden turns and asymmetries in the $\Delta \chi^2$ distributions of $E^{\rm tot}_{}$. These behaviors can be qualitatively explained by the boundary effect of $E^{\rm tot}_{\nu^{}_\alpha}$, which is considered to be non-negative in the physical range. When the deviation of $E^{\rm tot}_{}$ from its true value (i.e., $3\times 10^{53}$ erg) becomes larger, the best-fit values of $R^{}_{\nu^{}_{e}}$ in the NO case and of $R^{}_{\overline{\nu}_e}$ in the IO case will be zero in the marginalization process, resulting in sudden turns in the $\Delta \chi^2$ distributions of $E^{\rm tot}_{}$. Moreover, when one fits with a smaller $E^{\rm tot}_{}$, the scanned ranges of $E^{\rm tot}_{\nu^{}_\alpha}$ will accordingly become smaller to meet the non-negative requirements. If a larger $E^{\rm tot}_{}$ is going to be fitted, one has a wider range of $E^{\rm tot}_{\nu^{}_\alpha}$ to minimize the $\chi^2$ function. The asymmetric fitting process is an explanation for the observed asymmetries in the $\Delta \chi^2$ distributions. For both the NO and IO cases, the total energy $E^{\rm tot}_{}$ can be determined to be {$(3.0^{+0.42}_{-0.37})\times 10^{53}~{\rm erg}$ and at $(3.0^{+0.74}_{-0.33})\times 10^{53}~{\rm erg}$ the 1$\sigma$ C.L., respectively.} The precision ranges from {$15\%$ to $28\%$} depending on different combinations of detection processes.

\section{Concluding Remarks}

In this work, we have discussed the detection prospects of a galactic SN neutrino burst in the future large {liquid-}scintillator detectors. Taking the JUNO experiment as example, we have shown that a global analysis of different detection channels are important and complementary in constraining the spectral parameters of each neutrino species, testing the average-energy hierarchy of SN neutrinos and determining how the total energy is partitioned among neutrino flavors. When combined with the observations from large water- and ice-Cherenkov detectors, the multi-channel detection of galactic SN neutrinos in the {liquid-}scintillator detectors will be of particular importance to verify the neutrino-driven explosion mechanism of core-collapse supernovae.

First, ignoring neutrino flavor conversions, we investigate how well the average and total energies for each neutrino species, namely, $\overline{\nu}^{}_e$, $\nu^{}_e$ and $\nu^{}_x$ can be determined at the JUNO detector. Assuming that a total gravitational binding energy $3\times 10^{53}~{\rm erg}$ is equally distributed among all six neutrino and antineutrino flavors, we further take the flavor-dependent spectra in Eq.~(1), where the time-integrated neutrino energy spectra are parametrized by the average energy $\langle E^{}_\alpha \rangle$ and the spectral index $\gamma^{}_\alpha$. For a galatic SN at $D = 10~{\rm kpc}$, a mild hierarchy among average energies $(\langle E^{}_{\nu_e} \rangle, \langle E^{}_{\bar{\nu}_e} \rangle, \langle E^{}_{\nu_x} \rangle) = (12, 14, 16)~{\rm MeV}$ and a universal spectral index $\gamma^{}_\alpha = 3$, the numbers and energy distributions of neutrino events in different detection channels have been summarized in Table~\ref{table:events} and Fig. 1, based on the JUNO nominal setup.
Through a statistical analysis, we arrive at the following conclusions:
\begin{itemize}
\item For $\overline{\nu}^{}_e$, the precisions for the average energy $\langle E^{}_{\overline{\nu}_e} \rangle$ and the total energy $E^{\rm tot}_{\overline{\nu}_e}$ reach $1.4\%$ and $5.4\%$, respectively, at the $90\%$ C.L. Although the $^{12}{\rm B}$-CC channel also contributes, the precisions are dominated by the IBD channel due to a large number of events. The variation of the IBD detection efficiency from $90\%$ to $99\%$ does not change the results much, but it affects the other channels without coincident signals.

\item For $\nu^{}_e$, the average energy $\langle E^{}_{\nu_e} \rangle$ is mainly determined from the $e$ES+{$^{13}{\rm N}$-CC} and $^{12}{\rm C}$-CC channels, and the precision is found to be {$12\%$} at the $90\%$ C.L. On the other hand, a precision about {$27\%$} of the total energy $E^{\rm tot}_{\nu_e}$ can be reached by solely using the $e$ES-like signal, but it will be improved to {$24\%$} at the same C.L. when the constraint from the $^{12}{\rm C}$-CC is included.

\item For $\nu^{}_x$, the $p$ES, $e$ES+{$^{13}{\rm N}$-CC}, $^{12}{\rm C}$-NC {and $^{12}{\rm C}$-NC} are the most important channels. Regarding the average energy $\langle E^{}_{\nu_x} \rangle$, the precision is dominated by the large number of $p$ES events, and reaches {$5.2\%$} at the $90\%$ C.L. However, the total energy $E^{\rm tot}_{\nu_e}$ is most constrained by the $e$ES, as a $20\%$ uncertainty of the $p$ES cross section is assumed. The combined analysis of all relevant channels yields a precision of {$27\%$ for $E^{\rm tot}_{\nu_e}$ and $4.6\%$ for $\langle E^{}_{\nu_x} \rangle$} at the same C.L.
\end{itemize}
It is worth mentioning that the above conclusions depend on the assumption of the mild hierarchy among neutrino average energies. If higher average energies for $\nu^{}_e$ and $\nu^{}_x$ are taken, e.g., $(\langle E^{}_{\nu_e} \rangle, \langle E^{}_{\bar{\nu}_e} \rangle, \langle E^{}_{\nu_x} \rangle) = (12, 15, 18)~{\rm MeV}$, the precisions will be improved~\cite{Laha:2014yua}.

Second, motivated by the JUNO capabilities of detecting SN neutrinos of different flavors, we propose to test the hypothesis of energy equilibration, namely, $R^{}_{\nu_e} = R^{}_{\overline{\nu}_e} = R^{}_{\nu_x} = 1/3$, where the energy ratio for each neutrino species has been defined in Eq.~(9). Taking account of the MSW matter effects in the SN envelope, {we find at the $1\sigma$ level that $0 \lesssim R^{}_{\nu_e} \lesssim 0.60$, $0.23 \lesssim R^{}_{\overline{\nu}_e} \lesssim 0.49$ and $0.17 \lesssim R^{}_{\nu_x} \lesssim 0.57$ in the NO case, and $0.01 \lesssim R^{}_{\nu_e} \lesssim 0.63$, $0 \lesssim R^{}_{\overline{\nu}_e} \lesssim 0.78$ and $0.20 \lesssim R^{}_{\nu_x} \lesssim 0.48$ in the IO case. The total gravitational binding energy can be determined to be $(3.0^{+0.42}_{-0.37})\times 10^{53}~{\rm erg}$ with a precision of $13\%$ for NO, while $(3.0^{+0.74}_{-0.33})\times 10^{53}~{\rm erg}$ with a precision of $18\%$ for IO.}

Therefore, future large {liquid-}scintillator detectors JUNO, RENO-50 and LENA will offer us a great opportunity to detect galactic SN neutrinos in a number of different channels, which can be used to reconstruct average neutrino energies and the total gravitational binding energy. In addition, the real-time measurements in both {liquid-}scintillator and water-Cherenkov detectors will hopefully be able to distinguish the neutrino emission during three distinct stages, namely, the prompt $\nu^{}_e$ burst, accretion phase and cooling phase. All the information are indispensable for us to understand the dynamics of a core-collapse SN.

\section*{Acknowledgements}
The authors thank John Beacom for helpful discussions and invaluable suggestions. This work was supported in part by the National Natural Science Foundation of China under Grant Nos. 11135009 and 11305193, by the Strategic Priority Research Program of the Chinese Academy of Sciences under Grant No. XDA10010100, by the National Recruitment Program for Young Professionals and the CAS Center for Excellence in Particle Physics.

\section*{APPENDIX: Statistical analysis details}

In this appendix, we present a detailed description of our statistical analyses in this work. To analyze the detection ability for different neutrino flavors,
we assume the efficiencies of the IBD, the $p$ES and $e$ES processes as $95\%$, $99\%$, $99\%$ respectively, and those of ${^{12}}{\rm N}$-CC and ${^{12}}{\rm B}$-CC processes to be $90\%$.
We assume the efficiencies of other processes to be $100\%$ for simplicity. Different detection efficiencies of the IBD process are also considered to illustrate their quantitative effects.
Moreover, the ${^{12}}{\rm N}$-CC and ${^{12}}{\rm B}$-CC processes in Eqs.~(\ref{eq: CCnue}) and (\ref{eq: CCnueb}) are taken to be completely indistinguishable and thus combined to a single detection channel, denoted as ${^{12}}{\rm C}$-CC, in our studies. Similarly, we do not distinguish the electrons between the $e$ES and ${^{13}}{\rm N}$-CC processes and denote their combination as $e$ES+${^{13}}{\rm N}$-CC.
Next we discuss possible backgrounds for different channels. The un-tagged IBD events constitute the backgrounds of the $e$ES+${^{13}}{\rm N}$-CC, ${^{12}}{\rm C}$-NC and ${^{13}}{\rm C}$-NC processes. In addition, we consider the proton and electron discrimination using the pulse shape technique and assume a $1\%$ mis-identification probability, which will contribute additional backgrounds for the $e$ES+${^{13}}{\rm N}$-CC and $p$ES processes. In this study, the uncertainties on the signal efficiencies and backgrounds are neglected. A detailed simulation analysis on the detection efficiencies and backgrounds and the corresponding uncertainties will be presented elsewhere.

For systematics, we employ a $2\%$ detection uncertainty for all channels, and an additional $20\%$ cross-section uncertainty for the $p$ES, ${^{12}}{\rm C}$-NC, ${^{13}}{\rm N}$-CC and ${^{13}}{\rm C}$-NC processes. Besides these normalization uncertainties, an energy-related uncertainty is included in the $p$ES case by taking a $3\%$ relative uncertainty on the Birks' constant $k_{\rm B}$ of the proton quenching effect, where the central value of $k_{\rm B}$ is taken as $9.8\times 10^{-3}~{\rm cm}~{\rm MeV}^{-1}$~\cite{vonKrosigk:2013sa}. The quench effects of positrons or electrons~\cite{Zhang:2014iza} are neglected in the current study.

Next we are going to construct of the least-squares functions in our statistical analyses. In a particular detection process, we employ the Poisson-type $\chi^{2}$ distribution with proper pull terms taking account of the above experimental systematic factors:
\begin{eqnarray}
\chi^2_i = \sum^{N^i_{\rm bin}}_{j = 1} \left[ 2 (T^{}_{ij} - O^{}_{ij}) + 2 O^{}_{ij} \ln \frac{O^{}_{ij}}{T^{}_{ij}} \right]
+ \sum_k \frac{\epsilon^{2}_k}{\sigma^{2}_k} \; ,
\end{eqnarray}
where $i$ denotes the index of the reaction channels, $N^i_{\rm bin}$ the number of bins, $O^{}_{ij}$ the number of observed events in the $j$-th bin and $i$-th channel, $T^{}_{ij}$ the expected number of events. In practice, both the observed and expected events are divided into signals $s^{}_{ij}$ and backgrounds $b^{}_{ij}$, while the signals in the expected number of events are corrected as $\hat{s}^{}_{ij} = s^{}_{ij} (1 + \sum_k\alpha_{ik}\epsilon^{}_k)$ by taking into account possible systematic uncertainties. Here ${\sigma^{}_k}$ is the $k$-th systematic uncertainty, $\epsilon^{}_k$ is the corresponding nuisance parameter, and $\alpha_{ik}$ is the fraction of neutrino event contribution in the $i$-th energy bin for the $k$-th nuisance parameter.

In detection of the $\overline{\nu}^{}_e$ flavor neutrinos, we perform the statistical analysis of the IBD and ${^{12}}{\rm C}$-CC processes in the left panel of Fig.~\ref{fig:nuebar}.
For the $\nu^{}_e$ flavor neutrinos, the $e$ES + ${^{13}}{\rm N}$-CC and ${^{12}}{\rm C}$-CC processes are considered as the main detection channels as shown in the left panel of Fig.~\ref{fig:nue}. Finally, the main detection channels of the $\nu^{}_x$ flavor neutrinos in the left panel of Fig.~\ref{fig:nux} includes the $p$ES, $e$ES+${^{13}}{\rm N}$-CC, ${^{12}}{\rm C}$-NC and ${^{13}}{\rm C}$-NC processes. The efficiencies, backgrounds and systematics of different detection channels in the construction of all the above-mentioned $\chi^{2}$ functions are summarized in Table~\ref{table:chi2}.
\begin{table}[!t]
\centering
\begin{tabular}{ccccccccccccc}
\hline
\multicolumn{1}{c}{Detection channels} & \multicolumn{1}{c}{} & \multicolumn{1}{c}{$\nu$ Flavors} & \multicolumn{1}{c}{} & \multicolumn{1}{c}{Efficiency} & \multicolumn{1}{c}{} & \multicolumn{1}{c}{Backgrounds} & \multicolumn{1}{c}{} & \multicolumn{5}{c}{Systematics} \\
\hline
\multicolumn{1}{c}{IBD} & \multicolumn{1}{c}{} & \multicolumn{1}{c}{$\overline{\nu}_e$} & \multicolumn{1}{c}{} & \multicolumn{1}{c}{$95\%$} & \multicolumn{1}{c}{} & \multicolumn{1}{c}{None} & \multicolumn{1}{c}{} & \multicolumn{1}{l}{Detection} & \multicolumn{1}{c}{} & \multicolumn{1}{c}{} & \multicolumn{1}{c}{} & \multicolumn{1}{r}{$2\%$} \\
\hline
\multicolumn{1}{c}{${^{12}}{\rm C}$-CC} & \multicolumn{1}{c}{} & \multicolumn{1}{c}{$\overline{\nu}_e$ and ${\nu}_e$} & \multicolumn{1}{c}{} & \multicolumn{1}{c}{$90\%$} & \multicolumn{1}{c}{} & \multicolumn{1}{c}{None} & \multicolumn{1}{c}{} & \multicolumn{1}{l}{Detection} & \multicolumn{1}{c}{} & \multicolumn{1}{c}{} & \multicolumn{1}{c}{} & \multicolumn{1}{r}{$2\%$}\\
\hline
\multicolumn{1}{c}{\multirow {3}{*}{$p$ES}} & \multicolumn{1}{c}{} & \multicolumn{1}{c}{\multirow {3}{*}{$\overline{\nu}_e$, ${\nu}_e$ and ${\nu}_x$}} & \multicolumn{1}{c}{} &
\multicolumn{1}{c}{\multirow {3}{*}{$99\%$}} & \multicolumn{1}{c}{} & \multicolumn{1}{c}{\multirow {3}{*}{$e$ES}} & \multicolumn{1}{c}{} & \multicolumn{1}{l}{Detection} & \multicolumn{1}{c}{} & \multicolumn{1}{c}{} & \multicolumn{1}{c}{} & \multicolumn{1}{r}{$2\%$} \\
\multicolumn{1}{c}{} & \multicolumn{1}{c}{} & \multicolumn{1}{c}{} & \multicolumn{1}{c}{} & \multicolumn{1}{c}{} & \multicolumn{1}{c}{} & \multicolumn{1}{c}{} & \multicolumn{1}{c}{} & \multicolumn{1}{l}{Cross section} & \multicolumn{1}{c}{} & \multicolumn{1}{c}{} & \multicolumn{1}{c}{} & \multicolumn{1}{r}{$20\%$}\\
\multicolumn{1}{c}{} & \multicolumn{1}{c}{} & \multicolumn{1}{c}{} & \multicolumn{1}{c}{} & \multicolumn{1}{c}{} & \multicolumn{1}{c}{} & \multicolumn{1}{c}{} & \multicolumn{1}{c}{} & \multicolumn{1}{l}{$k_{\rm B}$} & \multicolumn{1}{c}{} & \multicolumn{1}{c}{} & \multicolumn{1}{c}{} & \multicolumn{1}{r}{$3\%$}\\
\hline
\multicolumn{1}{c}{$e$ES} & \multicolumn{1}{c}{} & \multicolumn{1}{c}{$\overline{\nu}_e$, ${\nu}_e$ and ${\nu}_x$} & \multicolumn{1}{c}{} & \multicolumn{1}{c}{$99\%$} & \multicolumn{1}{c}{} & \multicolumn{1}{c}{${^{13}}{\rm N}$-CC+IBD+$p$ES} & \multicolumn{1}{c}{} & \multicolumn{1}{l}{Detection} & \multicolumn{1}{c}{} & \multicolumn{1}{c}{} & \multicolumn{1}{c}{} & \multicolumn{1}{r}{$2\%$} \\
\hline
\multicolumn{1}{c}{\multirow {2}{*}{${^{13}}{\rm N}$-CC}} & \multicolumn{1}{c}{} & \multicolumn{1}{c}{\multirow {2}{*}{${\nu}_e$}} & \multicolumn{1}{c}{} & \multicolumn{1}{c}{\multirow {2}{*}{$100\%$}} & \multicolumn{1}{c}{} & \multicolumn{1}{c}{\multirow {2}{*}{$e$ES+IBD}} & \multicolumn{1}{c}{} & \multicolumn{1}{l}{Detection} & \multicolumn{1}{c}{} & \multicolumn{1}{c}{} & \multicolumn{1}{c}{} & \multicolumn{1}{r}{$2\%$}\\
\multicolumn{1}{c}{} & \multicolumn{1}{c}{} & \multicolumn{1}{c}{} & \multicolumn{1}{c}{} & \multicolumn{1}{c}{} & \multicolumn{1}{c}{} & \multicolumn{1}{c}{} & \multicolumn{1}{c}{} & \multicolumn{1}{l}{Cross section} & \multicolumn{1}{c}{} & \multicolumn{1}{c}{} & \multicolumn{1}{c}{} & \multicolumn{1}{r}{$20\%$}\\
\hline
\multicolumn{1}{c}{\multirow {2}{*}{${^{12}}{\rm C}$-NC}} & \multicolumn{1}{c}{} & \multicolumn{1}{c}{\multirow {2}{*}{$\overline{\nu}_e$, ${\nu}_e$ and ${\nu}_x$}} & \multicolumn{1}{c}{} & \multicolumn{1}{c}{\multirow {2}{*}{$100\%$}} & \multicolumn{1}{c}{} & \multicolumn{1}{c}{\multirow {2}{*}{$e$ES+IBD}} & \multicolumn{1}{c}{} & \multicolumn{1}{l}{Detection} & \multicolumn{1}{c}{} & \multicolumn{1}{c}{} & \multicolumn{1}{c}{} & \multicolumn{1}{r}{$2\%$} \\
\multicolumn{1}{c}{} & \multicolumn{1}{c}{} & \multicolumn{1}{c}{} & \multicolumn{1}{c}{} & \multicolumn{1}{c}{} & \multicolumn{1}{c}{} & \multicolumn{1}{c}{} & \multicolumn{1}{c}{} & \multicolumn{1}{l}{Cross section} & \multicolumn{1}{c}{} & \multicolumn{1}{c}{} & \multicolumn{1}{c}{} & \multicolumn{1}{r}{$20\%$}\\
\hline
\multicolumn{1}{c}{\multirow {2}{*}{${^{13}}{\rm C}$-NC}} & \multicolumn{1}{c}{} & \multicolumn{1}{c}{\multirow {2}{*}{$\overline{\nu}_e$, ${\nu}_e$ and ${\nu}_x$}} & \multicolumn{1}{c}{} & \multicolumn{1}{c}{\multirow {2}{*}{$100\%$}} & \multicolumn{1}{c}{} & \multicolumn{1}{c}{\multirow {2}{*}{$e$ES+IBD}} & \multicolumn{1}{c}{} & \multicolumn{1}{l}{Detection} & \multicolumn{1}{c}{} & \multicolumn{1}{c}{} & \multicolumn{1}{c}{} & \multicolumn{1}{r}{$2\%$} \\
\multicolumn{1}{c}{} & \multicolumn{1}{c}{} & \multicolumn{1}{c}{} & \multicolumn{1}{c}{} & \multicolumn{1}{c}{} & \multicolumn{1}{c}{} & \multicolumn{1}{c}{} & \multicolumn{1}{c}{} & \multicolumn{1}{l}{Cross section} & \multicolumn{1}{c}{} & \multicolumn{1}{c}{} & \multicolumn{1}{c}{} & \multicolumn{1}{r}{$20\%$}\\
\hline
\end{tabular}
\vspace{0.5cm}
\caption{Summary of the efficiencies, backgrounds and systematics of the detection channels considered in this work, where ${^{12}}{\rm C}$-CC
denotes a combination of the ${^{12}}{\rm N}$-CC and ${^{12}}{\rm B}$-CC processes. We also combine the $e$ES and ${^{13}}{\rm N}$-CC processes together
and denote them as $e$ES+${^{13}}{\rm N}$-CC.}
	\label{table:chi2}
\end{table}

When doing the global fitting of different reaction channels, we summate their respective $\chi^2$ functions by considering the full correlation of the detection systematic uncertainties and marginalize all the nuisance parameters in Table~\ref{table:chi2}. In the global analyses of Sec.~IV, we also marginalize other relevant SN neutrino flux parameters, i.e., $\langle E^{}_{\overline{\nu}_e}\rangle$, $\langle E^{}_{{\nu}_e}\rangle$, $\langle E^{}_{{\nu}_x}\rangle$ and $E^{\rm tot}_{}$ in Fig.~\ref{fig:ternary}, and $\langle E^{}_{\overline{\nu}_e}\rangle$, $\langle E^{}_{{\nu}_e}\rangle$, $\langle E^{}_{{\nu}_x}\rangle$ and $R^{}_{\nu^{}_{\alpha}}$ in Fig.~\ref{fig:lumi}.


\begin{thebibliography}{99}
\bibitem{Bethe:1990mw}
  H.~A.~Bethe,
  Rev.\ Mod.\ Phys.\  {\bf 62}, 801 (1990).

\bibitem{Wilson:1985}
  J.~R.~Wilson,
  in {\it Numerical Astrophysics}, edited by J. M. Centrella, J. M. LeBlanc, and R. L. Bowers (Jones \& Bartlett, Boston), p. 422.

\bibitem{Bethe:1984ux}
  H.~A.~Bethe and J.~R.~Wilson,
  Astrophys.\ J.\  {\bf 295}, 14 (1985).

\bibitem{Mirizzi:2015eza}
  A.~Mirizzi, I.~Tamborra, H.~T.~Janka, N.~Saviano, K.~Scholberg, R.~Bollig, L.~Hudepohl and S.~Chakraborty,
  Riv.\ Nuovo Cim.\  {\bf 39}, no. 1-2, 1 (2016)
  [arXiv:1508.00785].

\bibitem{Laha:2013hva}
  R.~Laha and J.~F.~Beacom,
  Phys.\ Rev.\ D {\bf 89}, no. 6, 063007 (2014)
  [arXiv:1311.6407].

\bibitem{Li:2013zyd}
  Y.~F.~Li, J.~Cao, Y.~Wang and L.~Zhan,
  Phys.\ Rev.\ D {\bf 88}, 013008 (2013)
  [arXiv:1303.6733].

\bibitem{An:2015jdp}
  F.~An {\it et al.} [JUNO Collaboration],
  J.\ Phys.\ G {\bf 43}, no. 3, 030401 (2016)
  [arXiv:1507.05613].

\bibitem{Lu:2014zma}
  J.~S.~Lu, J.~Cao, Y.~F.~Li and S.~Zhou,
  JCAP {\bf 05}, 044 (2015)
  [arXiv:1412.7418].

\bibitem{Rossi-Torres:2015rla}
  F.~Rossi-Torres, M.~M.~Guzzo and E.~Kemp,
  [arXiv:1501.00456].

\bibitem{Kim:2014rfa}
  S.~B.~Kim,
  Nucl.\ Part.\ Phys.\ Proc.\  {\bf 265-266}, 93 (2015)
  [arXiv:1412.2199].

\bibitem{Wurm:2011zn}
  M.~Wurm {\it et al.}  [LENA Collaboration],
  Astropart.\ Phys.\  {\bf 35}, 685 (2012)
  [arXiv:1104.5620].

\bibitem{Alimonti:2000xc}
  G.~Alimonti {\it et al.}  [Borexino Collaboration],
  Astropart.\ Phys.\  {\bf 16}, 205 (2002)
  [hep-ex/0012030].

\bibitem{Arafune:1988hx}
  J.~Arafune, M.~Fukugita, Y.~Kohyama and K.~Kubodera,
  Phys.\ Lett.\ B {\bf 217}, 186 (1989).

\bibitem{Fukugita:1989wv}
  M.~Fukugita, Y.~Kohyama, K.~Kubodera and T.~Kuramoto,
  Phys.\ Rev.\ C {\bf 41}, 1359 (1990).

\bibitem{Suzuki:2012aa}
  T.~Suzuki, A.~B.~Balantekin and T.~Kajino,
  Phys.\ Rev.\ C {\bf 86}, 015502 (2012)
  [arXiv:1204.4231].

\bibitem{Mollenberg:2014mfa}
  R.~M\"{o}llenberg, F.~von Feilitzsch, D.~Hellgartner, L.~Oberauer, M.~Tippmann, J.~Winter, M.~Wurm and V.~Zimmer,
  Phys.\ Lett.\ B {\bf 737}, 251 (2014)
  [arXiv:1408.0623].

\bibitem{Lujan-Peschard:2014lta}
  C.~Lujan-Peschard, G.~Pagliaroli and F.~Vissani,
  JCAP {\bf 1407}, 051 (2014)
  [arXiv:1402.6953].

\bibitem{Laha:2014yua}
  R.~Laha, J.~F.~Beacom and S.~K.~Agarwalla,
  [arXiv:1412.8425].

\bibitem{Keil:2002in}
  M.~T.~Keil, G.~G.~Raffelt and H.~-T.~Janka,
  Astrophys.\ J.\  {\bf 590}, 971 (2003)  [astro-ph/0208035].

\bibitem{Vogel:1999zy}
  P.~Vogel and J.~F.~Beacom,
  Phys.\ Rev.\ D {\bf 60}, 053003 (1999)  [hep-ph/9903554].

\bibitem{Strumia:2003zx}
  A.~Strumia and F.~Vissani,
  Phys.\ Lett.\ B {\bf 564}, 42 (2003)  [astro-ph/0302055].

\bibitem{Apollonio:1999jg}
  M.~Apollonio {\it et al.}  [CHOOZ Collaboration],
  Phys.\ Rev.\ D {\bf 61}, 012001 (2000).

\bibitem{Fischer:2015oma}
  V.~Fischer {\it et al.},
  JCAP {\bf 1508}, 032 (2015)
  [arXiv:1504.05466].

\bibitem{Beacom:2002hs}
  J.~F.~Beacom, W.~M.~Farr and P.~Vogel,
  Phys.\ Rev.\ D {\bf 66}, 033001 (2002)  [hep-ph/0205220].

\bibitem{Dasgupta:2011wg}
  B.~Dasgupta and J.~F.~Beacom,
  Phys.\ Rev.\ D {\bf 83}, 113006 (2011)  [arXiv:1103.2768].

\bibitem{vonKrosigk:2013sa}
  B.~von Krosigk, L.~Neumann, R.~Nolte, S.~Rottger and K.~Zuber,
  Eur.\ Phys.\ J.\ C {\bf 73}, 2390 (2013)
  [arXiv:1301.6403].

\bibitem{Zhang:2014iza}
  F.~H.~Zhang, B.~X.~Yu, W.~Hu, M.~S.~Yang, G.~F.~Cao, J.~Cao and L.~Zhou,
  Chin.\ Phys.\ C {\bf 39}, 016003 (2015)
  [arXiv:1403.3257].

\bibitem{Weinberg:1972tu}
  S.~Weinberg,
  Phys.\ Rev.\ D {\bf 5}, 1412 (1972).

\bibitem{Ahrens:1986xe}
  L.~A.~Ahrens {\it et al.},
  Phys.\ Rev.\ D {\bf 35}, 785 (1987).

\bibitem{Pagliaroli:2012hq}
  G.~Pagliaroli, C.~Lujan-Peschard, M.~Mitra and F.~Vissani,
  Phys.\ Rev.\ Lett.\  {\bf 111}, no. 2, 022001 (2013)
  [arXiv:1210.4225].

\bibitem{Fukugita:1988hg}
  M.~Fukugita, Y.~Kohyama and K.~Kubodera,
  Phys.\ Lett.\ B {\bf 212}, 139 (1988).

\bibitem{Volpe:2000zn}
  C.~Volpe, N.~Auerbach, G.~Colo, T.~Suzuki and N.~Van Giai,
  Phys.\ Rev.\ C {\bf 62}, 015501 (2000)  [nucl-th/0001050].

\bibitem{Auerbach:2001hz}
  L.~B.~Auerbach {\it et al.}  [LSND Collaboration],
  Phys.\ Rev.\ C {\bf 64}, 065501 (2001)  [hep-ex/0105068].

\bibitem{Marciano:2003eq}
  W.~J.~Marciano and Z.~Parsa,
  J.\ Phys.\ G {\bf 29}, 2629 (2003)  [hep-ph/0403168].


\bibitem{Hannestad:1999zy}
  S.~Hannestad, H.~T.~Janka, G.~G.~Raffelt and G.~Sigl,
  Phys.\ Rev.\ D {\bf 62}, 093021 (2000)
  [astro-ph/9912242].

\bibitem{Duan:2010bg}
  H.~Duan, G.~M.~Fuller and Y.~Z.~Qian,
  Ann.\ Rev.\ Nucl.\ Part.\ Sci.\  {\bf 60}, 569 (2010)
  [arXiv:1001.2799].

\bibitem{Mikheev:1986gs}
  S.~P.~Mikheev and A.~Y.~Smirnov,
  Sov.\ J.\ Nucl.\ Phys.\  {\bf 42}, 913 (1985)
  [Yad.\ Fiz.\  {\bf 42}, 1441 (1985)].

\bibitem{Mikheev:1986wj}
  S.~P.~Mikheev and A.~Y.~Smirnov,
  Nuovo Cim.\ C {\bf 9}, 17 (1986).

\bibitem{Wolfenstein:1977ue}
  L.~Wolfenstein,
  Phys.\ Rev.\ D {\bf 17}, 2369 (1978).

\bibitem{Dighe:1999bi}
  A.~S.~Dighe and A.~Y.~Smirnov,
  Phys.\ Rev.\ D {\bf 62}, 033007 (2000)
  [hep-ph/9907423].

\bibitem{Agashe:2014kda}
  K.~A.~Olive {\it et al.}  [Particle Data Group Collaboration],
  Chin.\ Phys.\ C {\bf 38}, 090001 (2014).

\end{thebibliography}
\end{document}